\documentclass[twocolumn]{elsarticle}

\usepackage{lineno,hyperref}
\usepackage{amsmath,amsfonts,amsthm,braket,graphicx}
\usepackage{dcolumn} %multicol
\journal{Computer Physics Communications}

\newcommand{\cnf}{\mathcal{R}}

\newcommand{\vett}[1]{{\bf{#1}}}

\newcommand{\parm}{{\bf{p}}}

\newcommand{\hefour}{ $^4\mbox{He}$ }

\begin{document}
%\nocite{*} % non stampare i bibitem non usati (Attenzione, attivare \usepackage{refcheck} per avere un warning)

\onecolumn
\begin{frontmatter}

%\twocolumn[
%  \begin{@twocolumnfalse}
\title{Implementation of the Linear Method 
       for the optimization of Jastrow-Feenberg and Backflow Correlations}

%\author{M. Motta\fnref{myfootnote}}
%\author{G. Bertaina\fnref{myfootnote}}
%\author{D. E. Galli\fnref{myfootnote}}

\author{M. Motta}
%\address{Universit\` a degli Studi di Milano, Via Celoria 16, 20133, Milano, Italy.}
\author{G. Bertaina}
%\address{Universit\` a degli Studi di Milano, Via Celoria 16, 20133, Milano, Italy.}
\author{D. E. Galli}
%\address{Universit\` a degli Studi di Milano, Via Celoria 16, 20133, Milano, Italy.}
\author{E. Vitali\corref{mycorrespondingauthor}}
\cortext[mycorrespondingauthor]{Corresponding author, Tel. +390250317664}
\address{Universit\` a degli Studi di Milano, Via Celoria 16, 20133, Milano, Italy.}
\ead{ettore.vitali@unimi.it}

%\author[mymainaddres]{M. Motta}
%\author[mymainaddres]{G. Bertaina}
%\author[mymainaddres]{D.E. Galli}
%\author[mymainaddres]{E. Vitali\corref{mycorrespondingauthor}}
%\ead{ettore.vitali@unimi.it, +390250317664}
%\address[mymainaddress]{Universit\` a degli Studi di Milano, Via Celoria 16, 20133, Milano, Italy.}

%\fntext[myfootnote]{Universit\` a degli Studi di Milano, Via Celoria 16, 20133, Milano, Italy.}

\begin{abstract}
We present a fully detailed and highly performing implementation of the Linear Method [J. Toulouse and C. J. Umrigar (2007), \cite{Toulouse2007}]  to optimize Jastrow-Feenberg 
and Backflow Correlations in many-body wave-functions, which are widely used in 
condensed matter physics. 
We show that it is possible to implement such optimization scheme performing analytical derivatives of the wave-function with respect to the variational parameters  achieving the best 
possible complexity $\mathcal{O}\left( N^3\right)$ in the number of particles $N$. 
\end{abstract}

\begin{keyword}
\texttt{Quantum Monte Carlo; Variational Monte Carlo; Optimization}

\end{keyword}

\end{frontmatter}

\twocolumn
\section{Introduction}
Within modern theoretical condensed matter physics, a very important role is played by
Wave-Function(WF) based methodologies \cite{Szabo1996,Whitlock1986}.
In particular, in the realm of Quantum Monte Carlo (QMC) techniques \cite{Whitlock1986} at 
zero temperature, accurate approximations of the ground state 
or of some excited states of the investigated system
are crucial. 
For simulations of Bose systems in their ground state, 
although projector ground state QMC methods have been shown 
to yield {\it{exact}}\cite{gfmc,rept,pigs,spigs,patate} results regardless 
of the employed trial wave-function, 
an accurate choice of latter improves the efficiency of the algorithm 
and provides a deep insight into the behavior of the system\cite{davide,ettore}.
On the other hand, for excited states of Bose systems and for Fermi systems,
the need of accurate WFs is a necessity stemming from the sign or
phase problem \cite{Feynman1965}.
Once given the Hamiltonian of a physical system, a functional form for the many-body wave-function
is typically guessed combining physical intuition and mathematical arguments based
on the imaginary time evolution\cite{Caffarel1988,Caffarel1989,Holz2003}.
In general, some parameters $\parm \in\mathcal{P}\subseteq\mathbb{R}^n$,
usually called variational parameters, remain to be determined.
One thus deals with a family of WFs:

\begin{equation}
\label{psiofp}
\parm \mapsto |\Psi(\parm)\rangle, \quad \langle \cnf  |\Psi(\parm)\rangle = \Psi(\parm,\cnf) 
%\Psi:\mathcal{P} \to \hs \quad\quad \parm \mapsto \Psi(\parm,\cnf)
\end{equation}
where $\cnf$ denotes the many-body configuration (possibly including spins) 
of the system.
An extremely important issue concerns the development
and implementation of efficient tools to find optimal parameters.
This aim is pursued choosing a suitable {\it{cost function}} to be
optimized, typically the
expectation value of the hamiltonian, the {\it{energy}}:

\begin{equation}
\label{energy}
\mathcal{E}(\parm) = 
\frac{ \braket{ \Psi(\parm) |\hat{H}| \Psi(\parm) } }
     { \braket{ \Psi(\parm) | \Psi(\parm) } }
\end{equation}
or the {\it{energy variance}} \cite{Umrigar1988}:

\begin{equation}
\label{variance}
\mathcal{S}(\parm) = 
\frac{ \braket{ \Psi(\parm) | \left(\hat{H} - \mathcal{E}(\parm) \right)^2 | \Psi(\parm) } }
     { \braket{ \Psi(\parm) | \Psi(\parm) } }
\end{equation}
If the number of parameters is large, systematic procedures to find out the minimum
have to be devised.
One of the most widely employed scheme to alter the variational
parameters is the \emph{correlated sampling}
(CS) method\cite{Whitlock1986}, in which a set of configurations distributed according to
$|\Psi(\parm_0,\cnf)|^2$ is generated, $\parm_0$ being the current parameter configuration.
With the purpose of minimizing the {\it{energy}}, such configurations are
used to estimate $\mathcal{E}(\parm)$ relying on the expression:
\begin{equation}
\mathcal{E}(\parm) = \frac{ \int d\cnf |\Psi(\parm_0,\cnf)|^2 \mathcal{W}(\cnf) E_L(\parm,\cnf) }
                          { \int d\cnf |\Psi(\parm_0,\cnf)|^2 \mathcal{W}(\cnf) }
\end{equation}
where: 
\begin{equation} 
\mathcal{W}(\cnf) = \frac{ |\Psi(\parm  ,\cnf)|^2 }
                         { |\Psi(\parm_0,\cnf)|^2 } \quad E_L(\parm,\cnf) = \frac{\hat{H}\Psi(\parm,\cnf)}{\Psi(\parm,\cnf)}
\end{equation}
The main advantage of the CS technique is that the sampling of $|\Psi(\parm_0,\cnf)|^2$
for a single parameter configuration $\parm_0$ gives access to the value of the
$\mathcal{E}(\parm)$, in principle, for any  parameter configuration $\parm$. 
$\mathcal{E}(\parm)$ is then minimized with respect to $\parm$ computing the energy gradient within
the forward difference approximation and updating $\parm$ with the Levenberg-Marquardt method
\cite{Levenberg1944,Marquardt1963}.

Although minimization of $\mathcal{E}(\parm)$ using the CS method has often been successful, in some
cases the procedure can exhibit a numerical instability\cite{Kent1999}: it is well known, in
particular, that the CS method may give inaccurate results when the nodal surface of a many-fermion
trial wave-function is allowed to change during the optimization process.
In fact, unless the nodal surfaces of $\Psi(\parm_0,\cnf)$ and $\Psi(\parm,\cnf)$ coincide, massive
fluctuations in the weights occur on configurations close to the zeros of $|\Psi(\parm_0,\cnf)|^2$,
determining drastic statistical errors in the CS estimate of $\mathcal{E}(\parm)$.

More recent optimization schemes\cite{Rappe2000,Rappe2005,Filippi2005,
Sorella2005,Umrigar2007,Huang1990,Nightingale2001,Toulouse2007} require
explicit calculations of derivatives of the form:

\begin{equation}
\label{problem}
 \left|\frac{\partial \Psi(\parm)}{\partial p_i}\right\rangle, \quad \hat{H}\,  \left|\frac{\partial \Psi(\parm)}{\partial p_i}\right\rangle
\end{equation}
with the aim of minimizing \eqref{energy} and/or \eqref{variance}.

Although \eqref{problem} are nothing but derivatives, their
na\"ive calculation and algorithmic implementation
leads, especially in the case of non-linear parameters,
to very computationally demanding optimization algorithms. 
It thus becomes necessary to devise
non trivial strategies to keep the complexity
of the calculations favorable.
In the present work we focus on a very
wide class of correlated many-body wave-functions,
very important for condensed matter physics: the
Slater-Jastrow-Three-body-Backflow (SJ3BBF) WF.
We show the possibility to compute  \eqref{problem}, for a given
variational parameter, performing
analytical derivatives, using at most $\mathcal{O}(N^3)$ operations,
$N$ being the number of particles. 
We provide a practical and fully detailed implementation of the Linear Method (LM), 
first conceived by Nightingale and Melik-Alaverdian \cite{Nightingale2001} and 
later generalized by Toulouse and Umrigar \cite{Toulouse2007} and Umrigar et
al. \cite{Umrigar2007} to the treatment of non-linear parameters.
In this paper we do not address the topic of the
scaling of the calculations with respect to the
number of variational parameters $M$, which is
discussed for example in the very interesting paper
\cite{sorella2010}

%Although \eqref{problem} are nothing but derivatives, their
%na\"ive calculation and algorithmic implementation leads, especially
%in the case of non-linear parameters,
%to very computationally demanding optimization algorithms.
%Practically, it is crucial to build up a strategy to control the
%scaling of the complexity with respect to the number of parameters $M$ and the
%number of particles $N$. In the present work we 
%focus on a very wide class of correlated many-body wave-functions, 
%very important for condensed matter physics: 
%the Slater-Jastrow-Three-body-Backflow (SJ3BBF) WF.
%We show the possibility to compute \eqref{problem}, 
%performing analytical derivatives, using $\mathcal{O}\left( M^2 \, N^3 \right)$ operations.
%We provide a practical and fully detailed implementation of the Linear Method (LM),
%first conceived by Nightingale and Melik-Alaverdian \cite{Nightingale2001}
%to optimize linear parameters (i.e. coefficients on which the wave-function $|\Psi(\parm) \rangle$ depends 
%inearly), 
% and later generalized by Toulouse and Umrigar \cite{Toulouse2007} and Umrigar et al.
%\cite{Umrigar2007} to the treatment of non-linear parameters.

\section{The Linear Method}
\label{sec:linear_method}

In order to keep a simple notation, we briefly describe here the LM
in the case of real-valued wave-functions.  The non trivial generalization
to the case of complex-valued
WFs is presented in \ref{appB}.
Within the LM, the optimization of the \emph{energy} \eqref{energy}
is pursued by iteratively:

\begin{enumerate}
\item expanding the normalized WF:
\begin{equation}
\label{lm1}
\ket{ \tilde{\Psi}(\parm) } = \frac{ \ket{\Psi(\parm) } }
                                   { \braket{\Psi(\parm)|\Psi(\parm)}^{\frac{1}{2}} }
\end{equation}
around the current parameter configuration $\parm_0$ to first order in the
parameter variation $\Delta \parm = \parm - \parm_0$:
\begin{equation}
\label{lm2}
\ket{ \overline{\Psi}(\parm) } = \ket{ \overline{\Psi}_0 } +
  \sum_{j=1}^{M} \Delta p_j \, \ket{ \overline{\Psi}_j }
\end{equation}
with $\ket{\overline{\Psi}_0} = \ket{ \overline{\Psi}(\parm_0) }$ and:

\begin{equation}
\label{lm4}
       \ket{ \overline{\Psi}_j } =
\frac{ \ket{ \Psi_j } }
     { \braket{\Psi_0 |\Psi_0 }^{\frac{1}{2} } }
 -
  \frac{ \braket{ \Psi_j | \Psi_0 } }
       { \braket{ \Psi_0 | \Psi_0 } }
  \, 
  \frac{ \ket{ \Psi_0 } }
       { \braket{ \Psi_0 | \Psi_0 }^{\frac{1}{2} } }
\end{equation}
where $\ket{ \Psi_0 } = \ket{ {\Psi}(\parm_0) } $ and
 $\ket{ \Psi_j } = \ket{ \frac{\partial \Psi}{\partial p_j}(\parm_0) }$.

The normalization constraint:
\begin{equation}
\label{lm5}
0 = \partial_{p_j} \braket{ \tilde{\Psi}(\parm) | \tilde{\Psi}(\parm)} =
                   2\braket{ \partial_{p_j} \tilde{\Psi}(\parm) | \tilde{\Psi}(\parm) }
%                 + \braket{ \overline{\Psi}(\parm) | \partial_{p_i} \overline{\Psi}(\parm) }
\end{equation}
results in the orthogonality between $\ket{\overline{\Psi}_0}$ and $\ket{\overline{\Psi}_i}$.
\item minimizing the expectation value of the Hamiltonian operator $\hat{H}$ over the WF
\eqref{lm2}:
\begin{equation}
\label{lm6}
\mathcal{E}(\parm) = 
\frac{ \braket{\overline{\Psi}(\parm)|\hat{H}|\overline{\Psi}(\parm)} }
     { \braket{\overline{\Psi}(\parm)|        \overline{\Psi}(\parm)} }
\end{equation}
with respect to the parameter variation $\Delta\parm$. Inserting \eqref{lm2} into \eqref{lm6} leads to:
\begin{equation}
\label{lm7}
\mathcal{E}(\parm) = 
\frac{ \begin{pmatrix} 1 \,\,  \Delta \parm^T \end{pmatrix}
       \begin{pmatrix} \mathcal{E}(\parm_0) & \vett{g}^T \\
                       \vett{g}          & \overline{\mathcal{H}} \\
       \end{pmatrix}
       \begin{pmatrix} 1 \\ \Delta \parm \end{pmatrix} }
     { \begin{pmatrix} 1 \,\,  \Delta \parm^T \end{pmatrix}
       \begin{pmatrix} 1                    & 0 \\
                       0           & \overline{\mathcal{S}} \\
       \end{pmatrix}
       \begin{pmatrix} 1 \\ \Delta \parm \end{pmatrix} }
\end{equation}
where $\mathcal{E}(\parm_0)$ is the current value of the energy,
$g_j=\frac{ \braket{\overline{\Psi}_0|\hat{H}|\overline{\Psi}_j} }
          { \braket{\overline{\Psi}_0|\overline{\Psi}_0} }$ is related to the gradient of the
energy by the following equality:
\begin{equation}
\partial_{p_j} \mathcal{E}(\parm_0) = 2g_j % + g^*_j
\end{equation}
which is easily derived computing $\partial_{p_i}\mathcal{E}(\parm)$ and recalling \eqref{lm5},
and $\overline{\mathcal{H}}_{ij} = \frac{ \braket{\overline{\Psi}_i|\hat{H}|\overline{\Psi}_j} }
                                        { \braket{\overline{\Psi}_0|\overline{\Psi}_0} }$.
%Similarly, $s_j=\frac{ \braket{\overline{\Psi}_0|\overline{\Psi}_j} }
%                     { \braket{\overline{\Psi}_0|\overline{\Psi}_0} }$ and
Similarly,
$\overline{\mathcal{S}}_{ij} = \frac{ \braket{\overline{\Psi}_i|\overline{\Psi}_j} }
                                    { \braket{\overline{\Psi}_0|\overline{\Psi}_0} }$.
%                                     For
 %     real-valued wave-functions, as discussed earlier, $s_j=0$.
      In published literature, the matrices appearing at the numerator and denominator of  \eqref{lm7}  are referred to, respectively, as  \emph{energy} and \emph{overlap} matrices \cite{Filippi2005,Umrigar2007,Nightingale2001}.

\item choosing the parameter variation $\Delta \parm$ in such a way to minimize \eqref{lm7}.
      The global minimum of \eqref{lm7} is necessarily a stationary point, where
      $\partial_\parm \mathcal{E}(\parm) = 0$; the stationarity condition translates
      into the following generalized eigenvalue equation\cite{nota_eigen}:
\begin{equation}
\label{lm8}
       \begin{pmatrix} \mathcal{E}(\parm_0) & \vett{g}^T \\
                       \vett{g}           & \overline{\mathcal{H}} \\
       \end{pmatrix}
       \begin{pmatrix} 1 \\ \Delta \parm \end{pmatrix} 
       =
       \mathcal{E}
       \begin{pmatrix} 1                    & 0 \\
                       0           & \overline{\mathcal{S}} \\
       \end{pmatrix}
       \begin{pmatrix} 1 \\ \Delta \parm \end{pmatrix}
\end{equation}
% RISOLVERE PER MEZZO DI 
      There are $M+1$ possible parameter variations $\{\Delta \parm^{(i)}\}_{i=1}^{M+1}$, 
      $M$ being the number of parameters, corresponding to properly rescaled solutions
      $\begin{pmatrix} 1 \\ \Delta \parm^{(i)} \end{pmatrix}$ of the generalized eigenvalue 
      equation \eqref{lm8} with eigenvalues $\{ \mathcal{E}^{(i)}\}_{i=1}^{M+1}$.
      Such parameter variations are stationary points of the energy expectation \eqref{lm7}.
      Moreover, inserting $\Delta \parm^{(i)}$ in \eqref{lm7} and recalling \eqref{lm8} leads
      to:
      \begin{equation}
      \label{phys_reas}
      \mathcal{E}(\parm_0+\Delta\parm^{(i)}) = \mathcal{E}^{(i)}
      \end{equation}
%      number of parameters, 
%      The eigenvectors $\{\Delta \parm_i \}_{i=1}^{M+1}$, $M$ being the
%      number of parameters,  {\color{blue} PER GIANLUCA, IL PROBLEMA AGLI
%      AUTOVALORI E' M+1 DIMENSIONALE, PROVIENE DA PSI0 E LE SUE M DERIVATE}
%      of \eqref{lm8} relative to the
%      eigenvalues $\{ \mathcal{E}_i \}_{i=1}^{M+1}$ are stationary points of the energy
%      expectation \eqref{lm7}.
%      Moreover, inserting $\Delta \parm_i$ in \eqref{lm7} and recalling \eqref{lm8} leads
%      to:
%      \begin{equation}
%      \mathcal{E}(\parm_0+\Delta\parm_i) = \mathcal{E}_i
%      \end{equation}
      clearly implying that the global minimum of the energy expectation \eqref{lm7} 
      corresponds to the solution of \eqref{lm8} relative to the lowest eigenvalue.
      It is worth noticing that, for large parameter variations $\Delta \parm^{(i)}$,
      the expanded WF \eqref{lm2} might not be an accurate approximation for the actual 
      normalized WF \eqref{lm1}. This could induce to unphysically low eigenvalues
      $\mathcal{E}^{(i)}$, that should be regarded to as unreliable estimates for the
      energy functional and rejected. 
\end{enumerate}

% Whitlock1986 M. H. Kalos and P. A. Whitlock, Monte Carlo Methods, Vol. 1 (Wiley-Interscience 1986).

%For instance, if the parameter set $\mathcal{P}$ is not closed under addition (a common situation
%for non-linear parameters), large parameter variations might take $\parm_0 + \Delta \parm$ ouside
%$\mathcal{P}$.

\subsection{VMC estimators of Energy and Overlap Matrices}

The elements of the energy and overlap matrices are estimated in
Variational Monte Carlo (VMC) calculations. Introducing the 
symbol $\left\langle f \right\rangle$ to denote the average:
\begin{equation}
\left\langle f \right\rangle = \int \, d\cnf \, \frac{ |\Psi_0(\cnf)|^2 }
                                          {\int \, d\cnf^\prime |\Psi_0(\cnf^\prime)|^2 } \, f(\cnf)
\end{equation}
of $f(\cnf)$ over the probability distribution 
                         $p (\cnf) = \frac{ |\Psi_0(\cnf)|^2 }
                                          {\int \, d\cnf^\prime |\Psi_0(\cnf^\prime)|^2 }$
evaluated using a large number of Monte Carlo configurations drawn from $p(\cnf)$.
It is readily found that:

%%%%%%%%%%%%%%%%%%%%%%%%%%%%%%%%%%%%%%%%%%%%%%%%%%%%%%%%%%%%%%%%%%%%%%%%%%%%%%%%%%%%%%%%%%%%
% per funzioni d'onda reali occorre considerare anche S_i0 e S_0i
% dobbiamo prendere una decisione!
% secondo me presentare l'estensione di un metodo senza applicazioni e' privo di senso
%%%%%%%%%%%%%%%%%%%%%%%%%%%%%%%%%%%%%%%%%%%%%%%%%%%%%%%%%%%%%%%%%%%%%%%%%%%%%%%%%%%%%%%%%%%%

\begin{equation}
\label{lm9}
\begin{split}
\overline{\mathcal{S}}_{ij} 
       =  \left\langle \frac{\Psi_i}{\Psi_0} \frac{\Psi_j}{\Psi_0} \right\rangle - 
          \left\langle \frac{\Psi_i}{\Psi_0} \right\rangle \left\langle \frac{\Psi_j}{\Psi_0} \right\rangle \\
\end{split}
\end{equation}
and that:
\begin{equation}
\label{lm10}
\begin{split}
\mathcal{E}(\parm_0)
       &= \left\langle E_L \right\rangle \\
g_j
       &= \left\langle E_{L,j} \right\rangle + \left\langle E_L \frac{\Psi_j}{\Psi_0} \right\rangle - 
         \left\langle E_L \right\rangle \left\langle \frac{\Psi_j}{\Psi_0} \right\rangle \\
g_i^T  
       &= \left\langle \frac{\Psi_i}{\Psi_0} E_L \right\rangle - \left\langle \frac{\Psi_i}{\Psi_0} \right\rangle \left\langle E_L \right\rangle \\
\overline{\mathcal{H}}_{ij}
       &= \left\langle \frac{\Psi_i}{\Psi_0} \frac{\Psi_j}{\Psi_0} E_L \right\rangle - 
         \left\langle \frac{\Psi_i}{\Psi_0} \right\rangle \left\langle \frac{\Psi_j}{\Psi_0} E_L \right\rangle  \\
        &-\left\langle \frac{\Psi_i}{\Psi_0} E_L \right\rangle \left\langle \frac{\Psi_j}{\Psi_0} \right\rangle + 
         \left\langle \frac{\Psi_i}{\Psi_0} \right\rangle \left\langle \frac{\Psi_j}{\Psi_0} \right\rangle \left\langle E_L \right\rangle \\
        &+\left\langle \frac{\Psi_i}{\Psi_0} E_{L,j} \right\rangle - \left\langle \frac{\Psi_i}{\Psi_0} \right\rangle \left\langle E_{L,j} \right\rangle 
\end{split}
\end{equation}
where the symbols
$E_L(\cnf) = \frac{\hat{H}\Psi_0(\cnf)}{\Psi_0(\cnf)}$ and
 $E_{L,j}(\cnf) = \frac{\hat{H}\Psi_j(\cnf)}{\Psi_0(\cnf)} - E_L(\cnf) \frac{\Psi_j(\cnf)}{\Psi_0(\cnf)}$
have been introduced.
 The estimators \eqref{lm9}, \eqref{lm10} are written in form of covariances rather than mean 
values of products to highlight their adequateness to numerical simulation, as it is a well 
known circumstance\cite{Filippi2005,Umrigar2007,Umrigar2008} that fluctuations of covariances 
are typically smaller than those of products.

%%%%%%%%%%%%%%%%%%%%%%%%%%%%%%%%%%%%%%%%%%%%%%%%%%%%%%%%%%%%%%%%%%%%%%%%%%%%%%%%%%%%%%%%%%%%%
% anche la discussione che segue e' per funzioni d'onda reali
%%%%%%%%%%%%%%%%%%%%%%%%%%%%%%%%%%%%%%%%%%%%%%%%%%%%%%%%%%%%%%%%%%%%%%%%%%%%%%%%%%%%%%%%%%%%%

The estimators for the elements $\mathcal{H}_{ij}$ of the energy matrix are not symmetric in $i$ and $j$ 
when approximated by averages over finite Monte Carlo samples, whereas $\mathcal{H}$ itself is of course
symmetric.   The hermiticity of the energy matrix is not exploited to symmetrize the estimator 
\eqref{lm9} since, as discussed in \cite{Nightingale2001,Umrigar2007}, using a non-symmetric 
estimator results in considerably smaller fluctuations  in the parameter variations than using 
its symmetrized analog.

We remark that, despite the solution of a non-symmetric eigenvalue equation can lead to complex
eigenvalues, it turns out \cite{Nightingale2001,Umrigar2007} that parameter variations $\Delta 
\parm_i$ corresponding to wave-functions $\Psi( \parm_0 + \Delta \parm_i )$ having large overlap
with the current wave-function $ \Psi( \parm_0) $ correspond to eigenvalues with small imaginary 
part.
Moreover, the leading divergences in  \eqref{lm10} near  the nodal surface of $\Psi_0$,
contained in the terms $\frac{\Psi_i(\cnf)}{\Psi_0(\cnf)}\frac{\Psi_j(\cnf)}{\Psi_0(\cnf)} E_L(\cnf)$ 
and $ \frac{\Psi_i(\cnf)} {\Psi_0(\cnf)} E_{L,j}(\cnf)$, cancel exactly \cite{Toulouse2007}, granting the adequateness of the linear method to the optimization of fermionic 
wave-functions.

\subsection{Alternative Normalization}
\label{sec:alternative_normalization}

The choice \eqref{lm1} is very natural but not unique. In fact, a differently normalized wave-function:
\begin{equation}
\label{lm12}
\ket{\tilde{\tilde{\Psi}}(\parm)} = N(\parm) \ket{\tilde{\Psi}(\parm)}
\end{equation}
has the first-order expansion:
\begin{equation}
\label{lm13}
\ket{\overline{\overline{\Psi}}(\parm)} = \ket{\overline{\Psi}_0} + 
                                          \sum_{j=1}^{M} \Delta p_j \ket{\overline{\overline{\Psi}}_j}
\end{equation}
under the condition that $N(\parm_0)=1$, with:
\begin{equation}
\ket{\overline{\overline{\Psi}}_j} = \ket{\overline{\Psi}_j} + \frac{\partial N}{\partial p_j}(\parm_0) \ket{\overline{\Psi}_0}
\end{equation}
The expansions \eqref{lm2} and \eqref{lm13} lie in the subspace of the Hilbert space spanned by the 
current wave-function $\ket{\Psi_0}$ and its derivatives $\ket{\Psi_j}$, implying that the parameter 
variations $ \Delta \parm $ and $ \Delta \overline{\overline{\parm }} $ corresponding to the energy 
minimum are proportional \cite{Umrigar2007}:
\begin{equation}
\label{eq:sorella1}
\Delta\overline{\overline{\parm}} = 
\frac{\Delta \parm}{1-\sum_{j=1}^{M} \frac{\partial N}{\partial p_j}(\parm_0) \Delta p_j}
\end{equation}
the derivatives $\frac{\partial N}{\partial p_j}(\parm_0)$ of the normalization function should be 
adjusted in such a way as to improve the performance of the algorithm. The empirical evidence that 
a good choice for nonlinear parameters is represented by:
\begin{equation}
\label{eq:sorella2}
\frac{ \partial N}{\partial p_j}(\parm_0) = - 
\frac{ (1-\xi) \sum_k \overline{\mathcal{S}}_{jk} \Delta p_k }
     { (1-\xi) + \xi \sqrt{1 + \sum_{jk} \Delta p_j \overline{\mathcal{S}}_{jk}} \Delta p_k }
\end{equation}
has been signaled in literature\cite{Sorella2001,Umrigar2007}. The constant $\xi\in [0,1]$ there
appearing is meant to be adjusted by hand during each iteration so that, to gain insight into the 
rationale behind its choice, it is worth inserting  \eqref{eq:sorella2} into  \eqref{eq:sorella1}
obtaining:
\begin{equation}
\label{eq:sorella3}
\Delta\overline{\overline{\parm}} =
\frac{\Delta \parm}{ 1 + \frac{ (1-\xi) Q }{ (1-\xi) + \xi \sqrt{ 1 + Q } }}
\end{equation}
where $Q = \sum_{jk} \Delta p_j \overline{\mathcal{S}}_{jk} \Delta p_k$ is a positive quantity,
the overlap matrix \eqref{lm9} being positive-definite since:
\begin{equation}
\overline{\mathcal{S}}_{jk} = 
\left\langle \left( \frac{\Psi_j}{\Psi_0} - \left\langle \frac{\Psi_j}{\Psi_0} \right\rangle \right)
             \left( \frac{\Psi_k}{\Psi_0} - \left\langle \frac{\Psi_k}{\Psi_0} \right\rangle \right)
\right\rangle
\end{equation}
In the light of this observation, the denominator appearing at the right member of \eqref{eq:sorella3} is a 
monotonically decreasing function of $\xi$ ranging from $1+Q$ to $1$, so that smaller values of $\xi$ decreases 
the parameter variations.
We remark that in some cases the choice $\xi=1$ can result in excessively large parameter variations that must 
be rejected; the safer choice $\xi = 0$, on the other hand, can lead to excessively small parameter variations 
that slow down the convergence of the algorithm.
The choice $\xi = \frac{1}{2}$ typically represents a good compromise between these two competing effects.

\subsection{Regularization}
\label{sec:regularization}

If the current parameter configuration $\parm_0$ is not sufficiently close to the minimum 
for the quadratic approximation of the energy to hold, or if the number of VMC samples employed 
to estimate the elements of the energy and overlap matrices is too small, 
and the latter are insufficiently accurate, the parameter variations $\Delta \parm^{(i)}$ 
proposed by the LM may give rise to worse updated wave-functions.
In such situation, it is convenient to apply a Tikhonov regularization 
\cite{Tikhonov1977,Filippi2005} to the energy matrix \eqref{lm8} by making the substitution: 
\begin{equation}
\label{lm11}
       \begin{pmatrix} \mathcal{E}(\parm_0) & \vett{g}^T \\
                       \vett{g}           & \overline{\mathcal{H}} \\
       \end{pmatrix}
\to 
       \begin{pmatrix} \mathcal{E}(\parm_0) & \vett{g}^T \\
                       \vett{g}           & \overline{\mathcal{H}} + \alpha \mathbb{I} \\
       \end{pmatrix}
\end{equation}
$\alpha$ being a real positive number, for large values of which the parameter variations 
$ \Delta \parm_i$ are easily shown to either diverge as 
$\Delta\parm_i = \alpha\vett{v} + \mathcal{O}(1)$, $\vett{v}$ being solution 
of the nonlinear system $\vett{v} =(\vett{g}\cdot\vett{v}) \, S\vett{v}$, or vanish as 
$\Delta\parm_i =\alpha^{-1} \vett{w} + \mathcal{O}
( \alpha^{-2} )$, being $ \vett{w} = - \vett{g}$. Therefore, vanishing parameter 
variations rotate from their original direction to the steepest descent direction in a nontrivial way.
The parameter $\alpha\in(0,\infty)$ is meant to be adjusted by hand before each iteration.
The criterion of choice is discussed in \cite{Toulouse2007}: for several values of
$\alpha$, the parameter variation associated to the lowest physically reasonable
eigenvalue \eqref{phys_reas} is used as an input to a VMC calculation; then, either the value
of $\alpha$ yielding the lowest VMC energy is chosen, or an interpolation is carried
out to identify the best value of $\alpha$.

\section{Application to Condensed Matter WFs}

Typical calculations in condensed matter physics involve wave-functions 
containing two-body and three-body correlations for Bose systems 
\cite{Kalos1980,Saverio1995}, and backflow correlations for Fermi systems  
\cite{Kalos1981,Carlson1989,Kwon1993,Saverio1995,Lopez2006}.  
In the case of Fermi system such trial WFs are not positive definite; 
nevertheless, it is well known  \cite{Ceperley1991,Foulkes2001} that in VMC calculations 
the Monte Carlo sampling can be restricted, without introducing any bias, to subsets of the configuration space where the sign of 
the trial wave-function is fixed, for instance positive.
Within such regions the trial wave-function can always be written in the form:
\begin{equation}
\begin{split}
\label{applica0}
\Psi(\parm,\cnf) &= e^{-Z(\parm,\cnf)} \\
Z(\parm,\cnf) &= 
\begin{cases}
Z_{2B}(\parm,\cnf)+Z_{3B}(\parm,\cnf)\phantom{+Z_{BF}(\parm,\cnf)}  \\
Z_{2B}(\parm,\cnf)+Z_{3B}(\parm,\cnf)         +Z_{BF}(\parm,\cnf)   \\
\end{cases} \\
\end{split}
\end{equation}
where the upper line refers to Bosons while the lower line to Fermions.

Explicitly, the two-body, three-body and backflow correlations 
have, quite generally, the following form:
\begin{equation}
\label{zeta}
\begin{split}
Z_{2B}(\parm,\cnf) &= \sum_{i<j=1}^N u(\parm_{2B},r_{ij}) \\
Z_{3B}(\parm,\cnf) &=  \frac{\lambda_T}{2} \sum_{l=1}^N \vett{G}_l(\parm_{3B},\cnf) \cdot 
                                                        \vett{G}_l(\parm_{3B},\cnf) \\ 
                   &-\lambda_T \sum_{i<j=1}^N \tilde{\xi}(\parm_{3B},r_{ij}) \\ 
Z_{BF}(\parm,\cnf) &= - \log\left( \det \left( \varphi_k( \vett{x}_i(\parm_{BF},\cnf) ) \right) \right) \\
\end{split}
\end{equation}
where the notation $\parm = (\parm_{2B},\lambda_T, \parm_{3B}, \parm_{BF})$ is used
to separate the variational parameters into subsets related to distinct WF parts.
In \eqref{zeta} $r_{ij} = |\vett{r}_i-\vett{r}_j|$ and:
\begin{equation}
\begin{split}
    \tilde{\xi}(\parm_{3B},r) &= \xi^2(\parm_{3B},r) r^2 \\
\vett{G}_l(\parm_{3B},\cnf) &= \sum_{i \neq l} \xi(\parm_{3B},r_{il}) \left( \vett{r}_l-\vett{r}_i \right) \\
\vett{x}_i(\parm_{BF},\cnf) &= \vett{r}_i + \sum_{j \neq i} \eta(\parm_{BF},r_{ij}) \left( \vett{r}_i-\vett{r}_j \right)
\end{split}
\end{equation}
for some parameter-dependent radial functions $u(\parm_{2B},r)$, $\xi(\parm_{3B},r)$ and $\eta(\parm_{BF},r)$. 
The VMC estimators for the energy and overlap matrices \eqref{lm9} contain the quantities \eqref{problem},
also occurring in the framework of other optimization techniques \cite{Rappe2000,Rappe2005,Filippi2005,
Sorella2005,Umrigar2007}, of which a completely explicit and numerically efficient expression will be 
now detailed.

First, we immediately observe that: 
\begin{equation}
\label{applica1}
\frac{\Psi_j(\parm,\cnf)}{\Psi(\parm,\cnf)} = 
-\partial_{p_j} \, Z(\parm,\cnf)
\end{equation}
is a sum of contributions, each of which is associated to a specific part of the wave-function. Moreover:
\begin{equation}
\label{applica2}
\begin{split}
& E_{L,j}(\parm,\cnf) = - \frac{\hbar^2}{2m} 
\sum_i \left( \frac{\triangle_i \Psi_j(\parm,\cnf)}{\Psi(\parm,\cnf)}  \right. \\
& - \left.               \frac{\Psi_j(\parm,\cnf)}{\Psi(\parm,\cnf)} 
\frac{\triangle_i \Psi(\parm,\cnf)}{\Psi(\parm,\cnf)} \right)
\end{split}
\end{equation}
and since:
\begin{equation}
\label{applica3}
\begin{split}
& - \frac{\triangle_i \Psi_j(\parm,\cnf)}{\Psi(\parm,\cnf)} = 
  \frac{\triangle_i \Psi(\parm,\cnf)}{\Psi(\parm,\cnf)}
  \partial_{p_j} Z(\parm,\cnf)  \\
& +
  \triangle_i \partial_{p_j} Z(\parm,\cnf)
+
2 \frac{\nabla_i \Psi(\parm,\cnf)}{\Psi(\parm,\cnf)} \cdot \nabla_i \partial_{p_j} Z(\parm,\cnf)
\end{split}
\end{equation}
merging \eqref{applica1} and \eqref{applica2} yields:
\begin{equation}
\label{applica4}
\begin{split}
& E_{L,j}(\parm,\cnf)= \frac{\hbar^2}{2m} 
\sum_i \left( \triangle_i \partial_{p_j} Z(\parm,\cnf)  \right. \\
& \left. + 2 \frac{\nabla_i \Psi(\parm,\cnf)}{\Psi(\parm,\cnf)} \cdot \nabla_i \partial_{p_j} Z(\parm,\cnf) \right) 
\end{split}
\end{equation}
Equations \eqref{applica1} and \eqref{applica4} pinpoint the need of computing the 
quantities $\partial_{p_j} Z(\parm,\cnf)$, $\nabla_i \partial_{p_j} Z(\parm,\cnf)$ 
and $\sum_i \triangle_i \partial_{p_j}   Z(\parm,\cnf)$ in order to construct the VMC 
estimators of the energy and overlap matrices.
We remark that, as the logarithm $-Z(\parm,\cnf)$ of the trial WF is additive in the 
terms associated to the many-body correlations it encompasses, its derivatives 
with respect to the variational parameters can be treated separately.    
In the forthcoming calculations, for all parameter-dependent radial functions 
$f(\parm,r)$ the notation  $\partial_{p_j} 
f^{(i)}(\parm,r)$ will be employed to indicate the $i$-th radial derivative of
$\partial_{p_j} f(\parm,r)$. Moreover, the symbol $r_{ij,\alpha}$ will be used
as a shortcut for $\left(\vett{r}_i-\vett{r}_j\right)_\alpha$ with $\alpha=1\dots d$,
$d$ being the dimensionality of the system.

\subsection{Two-Body Correlations}

The contribution to the quantity $Z(\parm,\cnf)$ brought by the two-body Jastrow factor reads:
\begin{equation}
\label{2b1}
Z(\parm,\cnf) = \sum_{k<l} u(\parm_{2B},r_{kl})
\end{equation}
so that:
\begin{equation}
\label{2b2}
\partial_{p_j} Z(\parm,\cnf) = \sum_{k<l} \partial_{p_j}u(\parm_{2B},r_{kl})
\end{equation}
and:
\begin{equation}
\partial_{r_{i\alpha}} \partial_{p_j} Z(\parm,\cnf) = \sum_{k\neq i} \partial_{p_j}u^{(1)}(\parm_{2B},r_{ik}) \frac{r_{ik,\alpha}}{r_{ik}}
\end{equation}
The laplacian $\sum_i \triangle_i \partial_{p_j} Z(\parm,\cnf)$ is readily obtained from:
\begin{equation}
\begin{split}
& \partial^2_{i,\alpha} \partial_{p_j} Z(\parm,\cnf) =
\sum_{k \neq i} \partial_{p_j}u^{(2)}(\parm_{2B},r_{ik}) \, \frac{ r^2_{ik,\alpha}  }{ r^2_{ik} }  \\
& +
         \frac{ \partial_{p_j}u^{(1)}(\parm_{2B},r_{ik}) }{ r_{ik}   } \, \frac{r^2_{ik} - r^2_{ik,\alpha}}{r^2_{ik}}
\end{split}
\end{equation}
and reads:
\begin{equation}
\label{2b4}
\begin{split}
& \sum_i \triangle_i \partial_{p_j} Z(\parm,\cnf) = 2 \left( \sum_{i<k} 
\partial_{p_j}u^{(2)}(\parm_{2B},r_{ik}) \right. \\
& \left. + (d-1) \, \frac{\partial_{p_j}u^{(1)}(\parm_{2B},r_{ik})}{r_{ik}} \right)
\end{split}
\end{equation}

\subsection{Backflow Correlations}

The contribution to the quantity $Z(\parm,\cnf)$ brought by the backflow correlations reads:
\begin{equation}
\label{bf1}
Z(\parm,\cnf) = - \log(\det(\varphi))
\end{equation}
where $\varphi_{ki} = \varphi_k(\parm_{BF},\vett{x}_i)$. In order to construct the VMC estimators of the 
energy and overlap matrices, the identities\cite{Magnus1999}:
\begin{eqnarray}
\label{bf2a}
\partial_{p_j} \det(\varphi) =   \det(\varphi) \, \mbox{tr}\left( \varphi^{-1} \partial_{p_j} \varphi \right) \\ \label{bf2b}
\partial_{r_{i\alpha}}  \varphi^{-1} = - \varphi^{-1} \, \partial_{r_{i\alpha}} \varphi \, \varphi^{-1} \\ \nonumber
\end{eqnarray}
will prove of fundamental importance. In fact:
\begin{equation}
\label{bf3}
\partial_{p_j} Z(\parm,\cnf) = - \mbox{tr}\left( \varphi^{-1} \partial_{p_j} \varphi \right)
\end{equation}
as immediate consequence of \eqref{bf2a}. Making use of \eqref{bf2b}, we readily obtain:
\begin{equation}
\label{bf4}
\begin{split}
&   \partial_{r_{i \alpha}} \partial_{p_j} Z(\parm,\cnf) = 
    \mbox{tr}\left( \varphi^{-1} \partial_{r_{i \alpha}} \varphi \, \varphi^{-1} \partial_{p_j} \varphi \right)  \\
& - \mbox{tr}\left( \varphi^{-1} (\partial_{r_{i \alpha}} \partial_{p_j} \varphi) \right)
\end{split}
\end{equation}
Eventually:
\begin{equation}
\label{bf5}
\begin{split}
\partial^2_{r_{i \alpha}} \partial_{p_j} Z(\parm,\cnf)&= \mbox{tr}\Big( \partial_{r_{i \alpha}} \left( \varphi^{-1} \partial_{r_{i \alpha}} \varphi \varphi^{-1} \right) \partial_{p_j} \varphi \Big)  \\
& 
                                                       + \mbox{tr}\Big( \left( \varphi^{-1} \partial_{r_{i \alpha}} \varphi \varphi^{-1} \right) (\partial_{r_{i \alpha}} \partial_{p_j} \varphi) \Big)  \\
                                                      &- \mbox{tr}\Big( \partial_{r_{i \alpha}} \varphi^{-1} \, \partial_{r_{i \alpha}} \partial_{p_j} \varphi \Big)  \\
&
                                                       - \mbox{tr}\Big( \varphi^{-1} \partial^2_{r_{i \alpha}} \partial_{p_j} \varphi \Big) \\
\end{split}
\end{equation}
Recalling \eqref{bf2a}, it is clear that the second and third terms of \eqref{bf5} 
are equal and opposite, implying that:
\begin{equation}
\label{bf6}
\begin{split}
\partial^2_{r_{i \alpha}} \partial_{p_j} Z(\parm,\cnf)&= \mbox{tr}\Big( \partial_{r_{i \alpha}} \left( \varphi^{-1} \partial_{r_{i \alpha}} \varphi \varphi^{-1} \right) \partial_{p_j} \varphi \Big)  \\
&                                                       + 2 \mbox{tr}\Big( \left( \varphi^{-1} \partial_{r_{i \alpha}} \varphi \varphi^{-1} \right) (\partial_{r_{i \alpha}} \partial_{p_j} \varphi) \Big)  \\
                                                      &- \mbox{tr}\Big( \varphi^{-1} \partial^2_{r_{i \alpha}} \partial_{p_j} \varphi \Big) \\
\end{split}
\end{equation}
Observing that:
\begin{equation}
\begin{split}
\partial_{r_{i \alpha}} \left( \varphi^{-1} \partial_{r_{i \alpha}} \varphi \varphi^{-1} \right) &= 
\varphi^{-1} \partial^2_{r_{i \alpha}} \varphi \varphi^{-1}  \\
& - 2 \varphi^{-1} \partial_{r_{i \alpha}} \varphi \varphi^{-1} \partial_{r_{i \alpha}} \varphi \varphi^{-1}
\end{split}
\end{equation} 
the following estimator for $\partial^2_{r_{i \alpha}} \partial_{p_j} Z(\parm,\cnf)$ is found:
\begin{equation}
\label{bf7}
\begin{split}
& \partial^2_{r_{i \alpha}} \partial_{p_j} Z(\parm,\cnf)= \mbox{tr}\Big( ( \varphi^{-1} \partial^2_{r_{i \alpha}} \varphi ) \varphi^{-1} \partial_{p_j} \varphi \Big)  \\
     &- 2 \mbox{tr}\Big( ( \varphi^{-1} \partial_{r_{i \alpha}} \varphi ) ( \varphi^{-1} \partial_{r_{i \alpha}} \varphi ) ( \varphi^{-1} \partial_{p_j} \varphi ) \Big) \\
                        &+ 2 \mbox{tr}\Big( ( \varphi^{-1} \partial_{r_{i \alpha}} \varphi ) ( \varphi^{-1} \partial_{r_{i \alpha}} \partial_{p_j} \varphi ) \Big) \\
& -   \mbox{tr}\Big( \varphi^{-1} \partial^2_{r_{i \alpha}} \partial_{p_j} \varphi \Big) \\
\end{split}
\end{equation}

Despite their slightly intricate appearance, the estimators \eqref{bf3}, \eqref{bf4} and \eqref{bf7} \emph{determine a total computational cost of the 
optimization procedure scaling as $\mathcal{O}(N^3)$}. This non-trivial result will be derived in detail in \ref{appA}.

\subsection{Three-Body Correlations}

The contribution to the quantity $Z(\parm,\cnf)$ brought by the three-body correlations reads:
\begin{equation}
\label{tb1}
Z(\parm,\cnf) = \frac{\lambda_T}{2} \sum_{l=1}^N \vett{G}(l) \cdot \vett{G}(l) - \lambda_T \sum_{j<k} \tilde{\xi}(r_{jk})
\end{equation}
If $p_j = \lambda_T$, the quantity $\partial_{p_j} Z(\parm,\cnf)$ is simply $\frac{1}{\lambda_T} \, Z(\parm,\cnf)$ so that:

\begin{equation}
\begin{split}
& \partial_{r_{i\alpha}} \partial_{p_j} Z(\parm,\cnf) = \\
& \sum_{l,\beta} \partial_{r_{i\alpha}} G_\beta(l) G_\beta(l) - \sum_{k\neq i} \frac{r_{ik,\alpha}}{r_{ik}}
\, \tilde{\xi}^{(1)}(r_{ik}) 
\end{split}
\end{equation}
and:
\begin{equation}
\begin{split}
& \sum_{\alpha} \partial^2_{r_{i\alpha}} \partial_{p_j} Z(\parm,\cnf) = \\
& 
\sum_{l,\alpha,\beta}
\left( \partial_{r_{i\alpha}} G_\beta(l) \partial_{i\alpha} G_\beta(l) +
       \partial^2_{r_{i\alpha}} G_\beta(l) G_\beta(l)
\right)
 \\
& -
\sum_{p \neq i} \left(\tilde{\xi}^{(2)}(r_{ip}) + (d-1) \frac{\tilde{\xi}^{(1)}(r_{ip})}{r_{ip}} \right) \\
\end{split}
\end{equation}
For all other parameters $p_j \in \parm_{3B}$:
\begin{equation}
\begin{split}
& \partial_{p_j} Z(\parm,\cnf) = \\
& \lambda_T \sum_{l,\beta} \partial_{p_j} G_\beta(l) G_\beta(l) - \lambda_T \sum_{l<k} \, \partial_{p_j} \tilde{\xi}(r_{lk})
\end{split}
\end{equation}
where:
\begin{equation}
\partial_{p_j} G_\beta(l) = 
\sum_{k\neq l} \partial_{p_j} \xi(r_{lk}) \, (\vett{r}_l - \vett{r}_k)
\end{equation}
Moreover:
\begin{equation}
\begin{split}
& \partial_{r_{i \alpha}} \partial_{p_j} Z(\parm,\cnf) = \\
& \lambda_T \sum_{l,\beta} \left( \partial_{r_{i \alpha}} \partial_{p_j} G_\beta(l) G_\beta(l) + 
\partial_{r_{i \alpha}} G_\beta(l) \partial_{p_j} G_\beta(l) \right)  \\
&- \lambda_T \sum_{k\neq i} \frac{r_{ik,\alpha}}{r_{ik}} \, \partial_p \tilde{\xi}^{(1)}(r_{ik})  \\
\end{split}
\end{equation}
with:
\begin{equation}
\begin{split}
& \partial_{r_{i \alpha}} \partial_{p_j} G_\beta(l) = \\
&= \delta_{li} 
\sum_{p \neq l} 
\left( 
\delta_{\alpha\beta} \partial_{p_j} \xi(r_{lp}) + \frac{\partial_{p_j} \xi^{(1)}(r_{lp})}{r_{lp}} \, r_{lp,\alpha} r_{lp,\beta} 
\right)  \\ 
&- (1-\delta_{li}) 
\left( 
\delta_{\alpha\beta} \partial_{p_j} \xi(r_{li}) + \frac{\partial_{p_j} \xi^{(1)}(r_{li})}{r_{li}} \, r_{li,\alpha} r_{li,\beta} 
\right)
\end{split}
\end{equation}
%with: 
%\begin{equation}
%\begin{split}
%& \partial_p \tilde{\xi}^{(1)}(r) = 
%2 \xi'(r) \partial_{p_j} \xi(r) r^2 + \\
%& + 2 \xi(r) \left( \partial_{p_j} \xi'(r) r^2 + \partial_{p_j} \xi(r) 2 r \right)
%\end{split}
%\end{equation}
The only remaining quantity is: 
\begin{equation}
\begin{split}
& \sum_{\alpha} \partial^2_{r_{i \alpha}} \partial_{p_j} Z(\parm,\cnf) = \\
& \lambda_T
\sum_{l,\alpha\beta}
\Big( 2 \partial_{r_{i \alpha}} \partial_{p_j} G_\beta(l) \partial_{r_{i \alpha}} G_\beta(l) \\
& + 
\partial^2_{r_{i \alpha}} \partial_{p_j} G_\beta(l) G_\beta(l) +
\partial^2_{r_{i \alpha}} G_\beta(l) \partial_{p_j} G_\beta(l) \Big)
\\
- & \lambda_T
\sum_{p \neq i} \left( \partial_{p_j} \tilde{\xi}^{(2)}(r_{ip}) + 
(d-1) \frac{\partial_{p_j}\tilde{\xi}^{(1)}(r_{ip})}{r_{ip}} \right) \\
\end{split}
\end{equation}
with:
\begin{equation}
\begin{split}
&\partial_{p_j} \sum_\alpha \partial^2_{r_{i\alpha}} G_\beta(l) = \\
&\delta_{il} \sum_{p\neq l} 
\left(
(d+1) \partial_{p_j} \xi^{(1)}(r_{lp}) \frac{r_{lp,\beta}}{r_{lp}} + \partial_{p_j} \xi^{(2)}(r_{lp}) r_{lp,\beta}
\right)
\\
+&(1-\delta_{il})
\left(
(d+1) \partial_{p_j} \xi^{(1)}(r_{li})\frac{r_{li,\beta}}{r_{li}} + \partial_{p_j} \xi^{(2)}(r_{li}) r_{li,\beta}
\right) \\
\end{split}
\end{equation}

\section{Results}

The performance of the algorithm has been benchmarked simulating a $3D$ system of $N=64$ 
$^4 \mbox{He}$ atoms
interacting through the HFDHE2 potential \cite{Aziz1979} near the equilibrium density 
$n=0.02186\,\mbox{\AA}^{-3}$, by 
making use of a wave-function encompassing Jastrow-McMillan two-body correlations \cite{McMillan1964}:
%%%%%%%%%%%%%%%%%%%%%%%%%%%%%%%%%%%%%%%%%%%%%%%%%%%%%
% unita' di misura densita' equilibrio elio liquido %
%%%%%%%%%%%%%%%%%%%%%%%%%%%%%%%%%%%%%%%%%%%%%%%%%%%%%
\begin{equation}
u(r) = \frac{1}{2} \left( \frac{b}{r} \right)^m
\end{equation}
and gaussian three-body correlations \cite{Carlson1989,Saverio1995}:
\begin{equation}
\xi(r) = e^{- w_T^2 \, (r-r_T)^2 }
\end{equation}
and a $2D$ system of $N=26$ electrons at Wigner-Seitz radius $r_s = 1$, by making use of a 
wave-function encompassing:
\begin{enumerate}
\item parameter-free RPA two-body correlations \cite{Gaskell1961,Gaskell1962,Ceperley1989,Kwon1993}:
\begin{equation}
2 n u(k) =  \sqrt{ \frac{1}{S_0^2(k)} + \frac{4 v(k) m n}{\hbar^2 k^2} } - \frac{1}{S_0(k)} 
\end{equation}
here detailed in Fourier space with $v(k) = \frac{2\pi e^2}{|k|}$ and:
\begin{equation}
\frac{\pi}{2} S_0(k) = \mbox{sin}^{-1}\left(\frac{k}
{2 k_F} \right) + \frac{k}{2 k_F} \sqrt{1-\left(\frac{k}{2 k_F}\right)^2 }
\end{equation}
\item rational backflow correlations parametrized as in \cite{Kwon1993}:
\begin{equation}
\eta(r) = \lambda_B \frac{1 + s_B r}{r_B + w_B r + r^{\frac{7}{2}}}
\end{equation}
\item and gaussian three-body correlations \cite{Carlson1989,Kwon1993,Saverio1995}:
\begin{equation}
\xi(r) = e^{- w_T^2 \, (r-r_T)^2 }
\end{equation}
\end{enumerate}

\subsection{The case of \texorpdfstring{$^4\text{He}$}{4He}}

%%%%%%%%%%%%%%%%%%%%%%%%%%%%%%%%%%%%%%%%%%%%%%%%%%%%%%%%%%%%%%%%%%%%%%%%%%%%%%%%%%%%%%%%%%%%%%%%%%%%%%%%%%%%%%%%%%%%%%%%%%%%%%%%
% ANDAMENTO ENERGIA VMC HE4 2B
\begin{figure}
\includegraphics[width=0.5\textwidth]{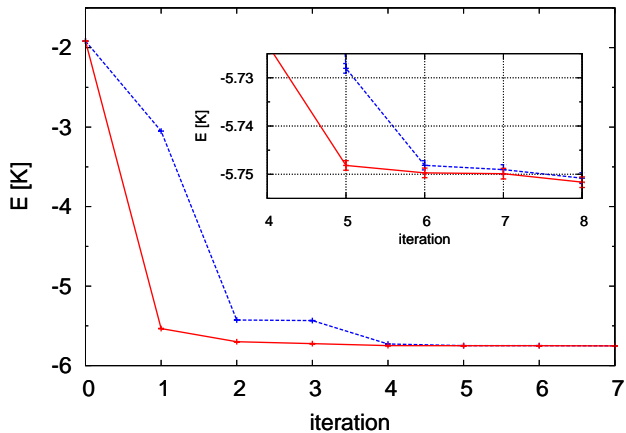}
\caption{(color online) Convergence of the VMC total energy of $N=64$ \hefour atoms at equilibrium
density during the optimization of the Jastrow-McMillan factor with (solid) and without
(dashed) regularization.}
\label{fig:1}
\end{figure}
% ANDAMENTO PARAMETRI HE4 2B
\begin{figure}
\includegraphics[width=0.5\textwidth]{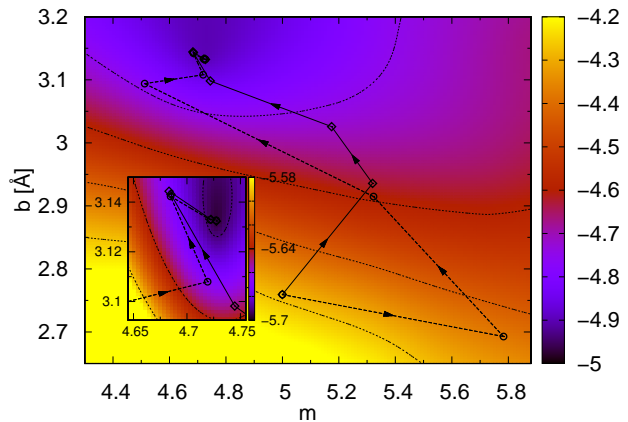}
\caption{(color online) Convergence of the Jastrow-McMillan factor of $N=64$ \hefour atoms at 
equilibrium density with (solid arrows) and without (dashed arrows) regularization.
To show the improvement brought by the regularization, the flows in the parameter space impressed 
by the two algorithms are superposed to a contour plot of the energy landscape $\mathcal{E}(\parm)$
obtained via several VMC calculations; dotted lines represent level curves of the energy landscape.}
\label{fig:2}
\end{figure}
%%%%%%%%%%%%%%%%%%%%%%%%%%%%%%%%%%%%%%%%%%%%%%%%%%%%%%%%%%%%%%%%%%%%%%%%%%%%%%%%%%%%%%%%%%%%%%%%%%%%%%%%%%%%%%%%%%%%%%%%%%%%%%%%
The Jastrow-McMillan factor has been first optimized in absence of three-body correlations: the convergence of the VMC energy is 
illustrated in figure \eqref{fig:1},  and the flow in the parameter space impressed by the optimization algorithm is illustrated
in figure \eqref{fig:2}. In both figures, two distinct series have been obtained by applying the basic parameter update algorithm described in section \eqref{sec:linear_method}, and by improving it with the alternative normalization and    
regularization procedures illustrated in subsections \eqref{sec:alternative_normalization} and \eqref{sec:regularization}
respectively.
% dashed lines have been obtained by applying the basic parameter update algorithm 
%described in section \eqref{sec:linear_method}, and solid lines by improving it with the alternative normalization and 
%regularization procedures illustrated in subsections \eqref{sec:alternative_normalization} and \eqref{sec:regularization} 
%respectively.
Figures \eqref{fig:1} and \eqref{fig:2} show that the use of alternative normalization and regularization results in a more 
rapid convergence of the algorithm.
We obtain an energy $-5.752(1) \, K$, in good agreement with the value $-5.72(2) \, K$ reported in \cite{Ceperley1996}.
%%%%%%%%%%%%%%%%%%%%%%%%%%%%%%%%%%%%%%%%%%%%%%%%%%%%%%%%%%%%%%%%%%%%%%%%%%%%%%%%%%%%%%%%%%%%%%%%%%%%%%%%%%%%%%%%%%%%%%%%%%%%%%%%
% ANDAMENTO ENERGIA VMC HE4 3B
\begin{figure}
\includegraphics[width=0.5\textwidth]{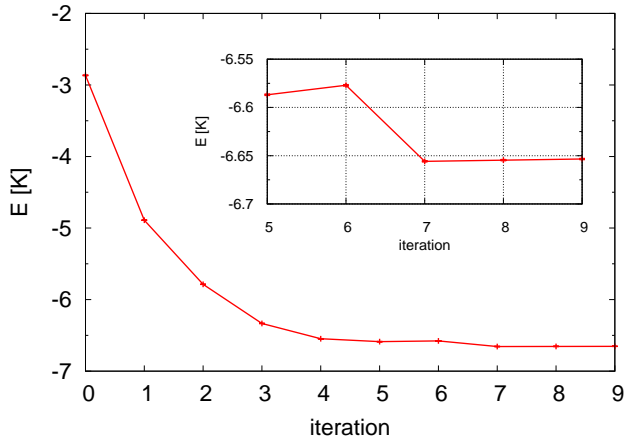}
\caption{(color online) Convergence of the VMC total energy of $N=64$ \hefour atoms at equilibrium
density during the optimization of the gaussian three-body factor in a wave function composed of a
three-body part multiplied by a previously-optimized Jastrow-McMillan factor.}
\label{fig:3}
\end{figure}
% ANDAMENTO PARAMETRI HE4 3B
\begin{figure}
%\begin{center}
\includegraphics[width=0.5\textwidth]{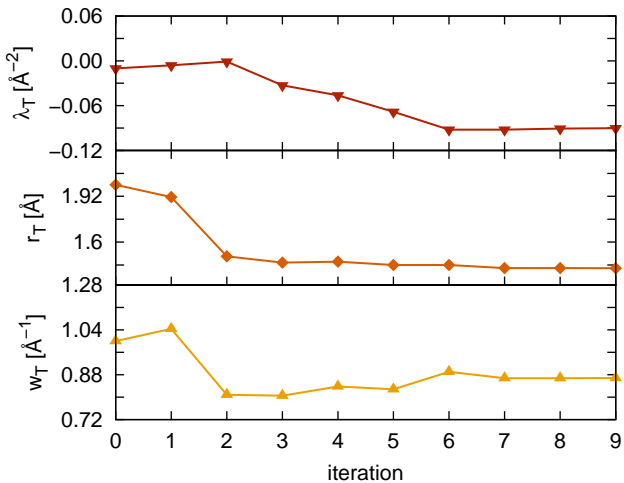}
\caption{Convergence of the gaussian three-body factor of $N=64$ \hefour atoms at equilibrium density.}
%\end{center}
\label{fig:4}
\end{figure}
%%%%%%%%%%%%%%%%%%%%%%%%%%%%%%%%%%%%%%%%%%%%%%%%%%%%%%%%%%%%%%%%%%%%%%%%%%%%%%%%%%%%%%%%%%%%%%%%%%%%%%%%%%%%%%%%%%%%%%%%%%%%%%%%
The gaussian factor has been subsequently optimized keeping the Jastrow-McMillan factor fixed 
at the parameter values $\parm_{2B}=(b,m)$ corresponding to the last step of figure \eqref{fig:2}.        The convergence of the VMC energy is illustrated in figure 
\eqref{fig:3}, and the flow in the parameter space in figure \eqref{fig:4}.
%We obtain an energy $-6.675(1)$, in good agreement with the value $-6.65(2) K$ reported in 
%\cite{Ceperley1996}.
%%%%%%%%%%%%%%%%%%%%%%%%%%%%%%%%%%%%%%%%%%%%%%%%%%%%%%%%%%%%%%%%%%%%%%%%%%%%%%%%%%%%%%%%%%%%%%%%%%%%%%%%%%%%%%%%%%%%%%%%%%%%%%%%
% ANDAMENTO ENERGIA VMC HE4 TOTALE
\begin{figure}
\includegraphics[width=0.5\textwidth]{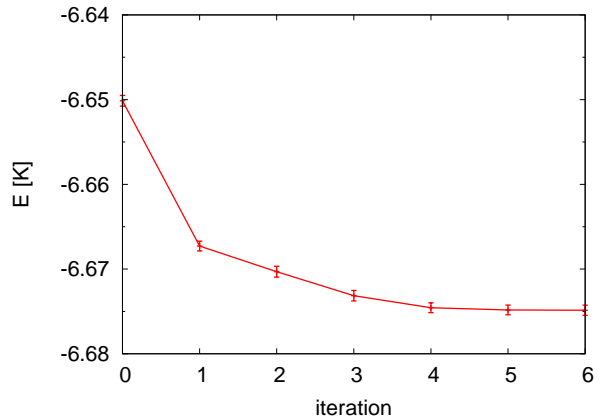}
\caption{(color online) Convergence of the VMC total energy of $N=64$ \hefour atoms at equilibrium
density during the optimization of a wave function composed of a Jastrow-McMillan factor and of a
gaussian three-body factor.}
\label{fig:5}
\end{figure}
% ANDAMENTO PARAMETRI HE4 TOTALE
\begin{figure}
%\begin{center}
\includegraphics[width=0.5\textwidth]{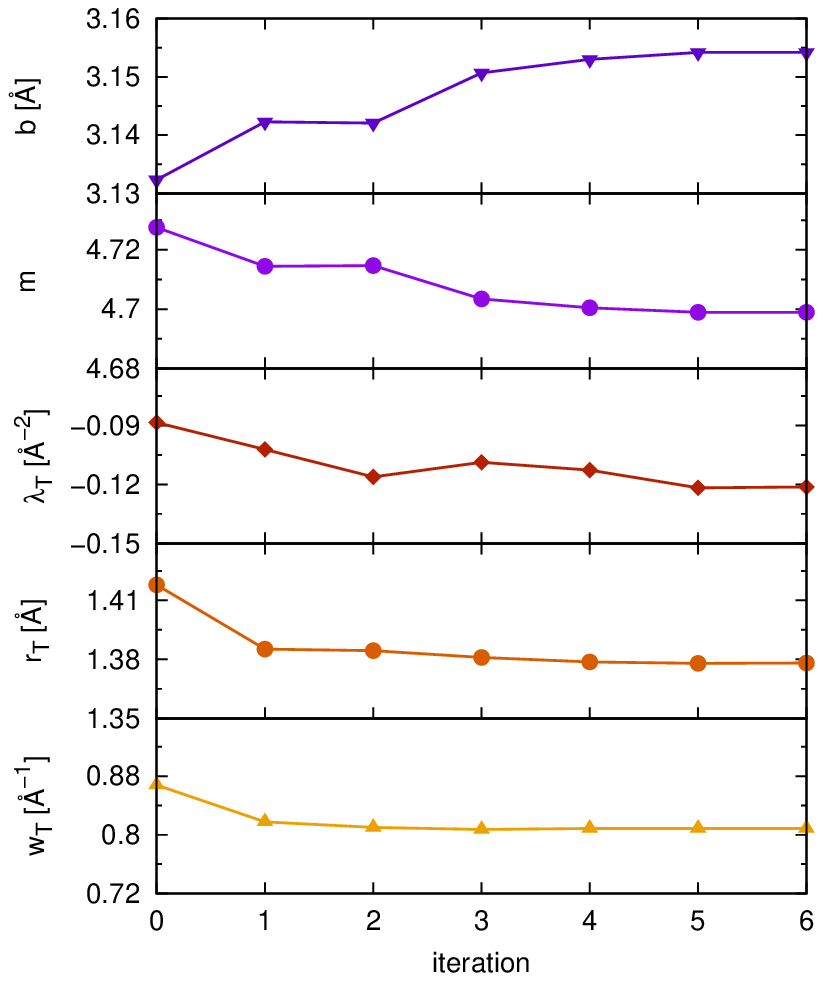}
$ $
\caption{Convergence of Jastrow-McMillan (upper panels) and gaussian three-body (lower panels) factors
of $N=64$ \hefour atoms at equilibrium density.}
%\end{center}
\label{fig:6}
\end{figure}
%%%%%%%%%%%%%%%%%%%%%%%%%%%%%%%%%%%%%%%%%%%%%%%%%%%%%%%%%%%%%%%%%%%%%%%%%%%%%%%%%%%%%%%%%%%%%%%%%%%%%%%%%%%%%%%%%%%%%%%%%%%%%%%%
A simultaneous optimization of the two-body and three-body correlations has been finally carried out, starting from the parameter values corresponding to the
last step of figures \eqref{fig:2} and \eqref{fig:4},  leading to the 
results illustrated in figures \eqref{fig:5} and \eqref{fig:6}.
We obtain an energy $-6.675(1) \, K$, in good agreement with the value $-6.65(2) \, K$ reported in 
\cite{Ceperley1996}.

\subsection{The case of \texorpdfstring{$2D$}{2D} electrons}

\begin{figure}
\includegraphics[width=0.5\textwidth]{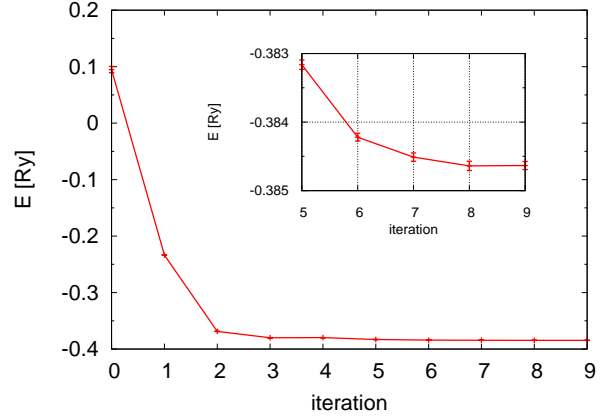}
\caption{(color online) Convergence of the VMC total energy of a 2D system of $N=26$ electrons
at $r_s = 1$ during the optimization of the backflow correlations of a wave-function composed of
a parameter-free Jastrow-RPA factor and a Slater-backflow determinant.}
\label{fig:7}
\end{figure}

\begin{figure}
\includegraphics[width=0.5\textwidth]{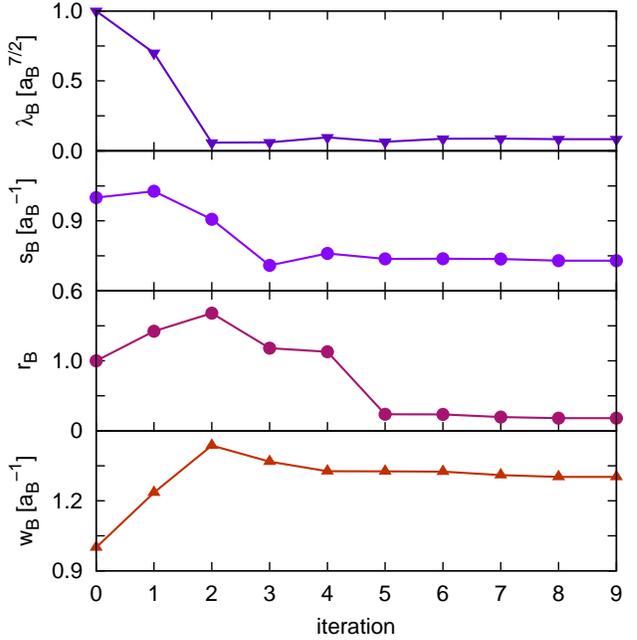}
\caption{(color online) Convergence of backflow correlations of a 2D system of $N=26$ electrons
at $r_s = 1$ during the optimization of the backflow correlations of a wave-function composed of
a parameter-free Jastrow-RPA factor and a Slater-backflow determinant.}
\label{fig:8}
\end{figure}

\begin{figure}
\includegraphics[width=0.5\textwidth]{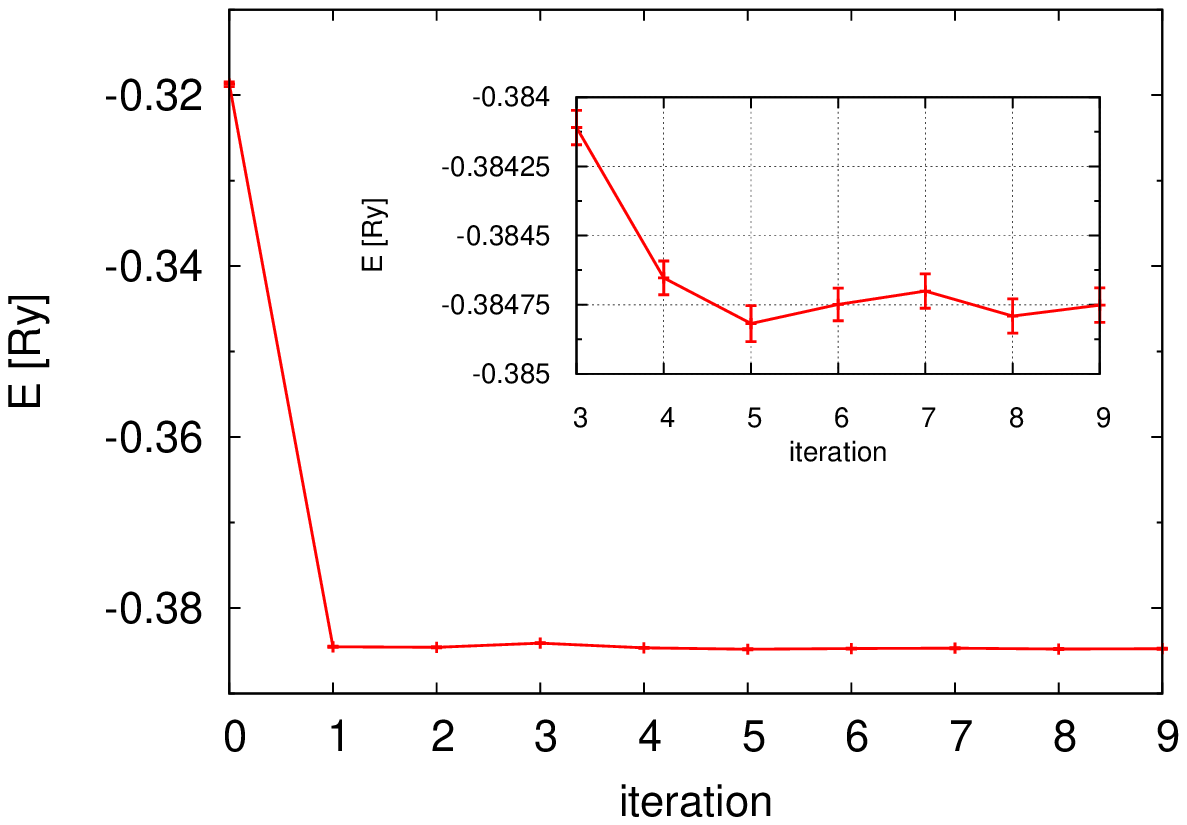}
\caption{(color online) Convergence of the VMC total energy of a 2D system of $N=26$ electrons
at $r_s =1$ during the optimization of a gaussian three-body factor of a wave-function composed
of a parameter-free Jastrow-RPA factor, a previously-optimized Slater-backflow determinant and
a gaussian three-body factor.}
\label{fig:9}
\end{figure}

\begin{figure}
\includegraphics[width=0.5\textwidth]{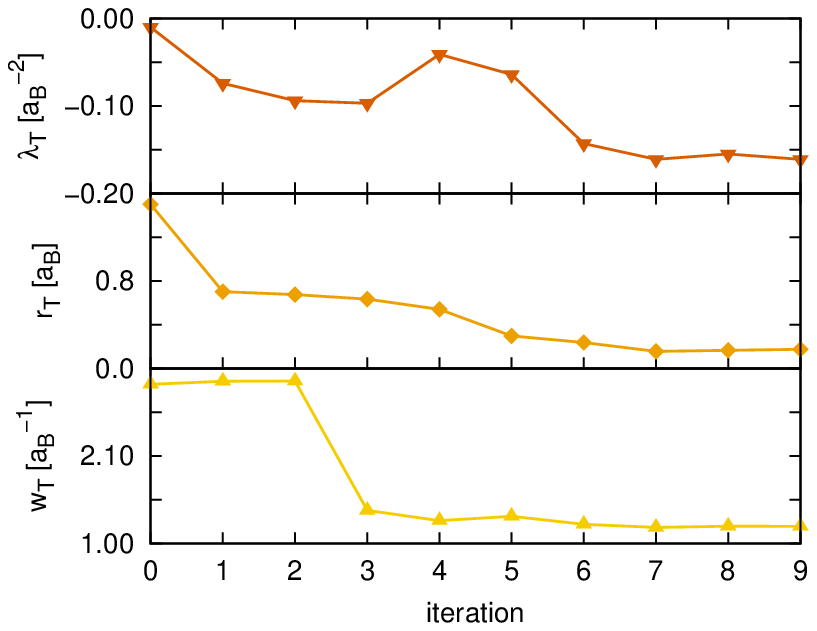}
\caption{(color online) Convergence of the gaussian three-body factor of a 2D system of $N=26$
electrons at $r_s =  1$ during the optimization of the gaussian three-body factor of a wave-function
composed of a parameter-free Jastrow-RPA factor, a previously-optimized Slater-backflow determinant
and a gaussian three-body factor.}
\label{fig:10}
\end{figure}

\begin{figure}
\includegraphics[width=0.5\textwidth]{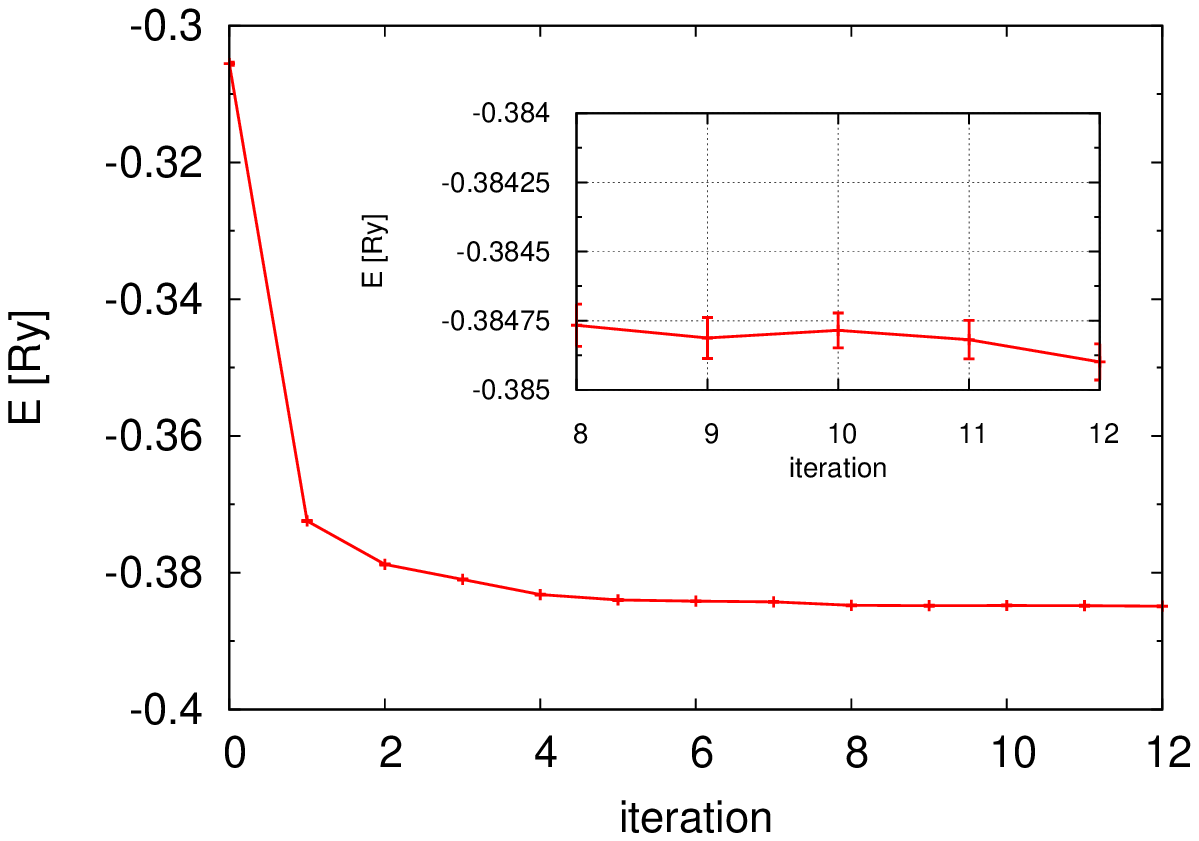}
\caption{(color online) Convergence of the VMC total energy of a 2D system of $N=26$ electrons
at $r_s =1$ during the simultaneous optimization of backflow and three-body correlations.}
\label{fig:11}
\end{figure}

\begin{figure}
\includegraphics[width=0.5\textwidth]{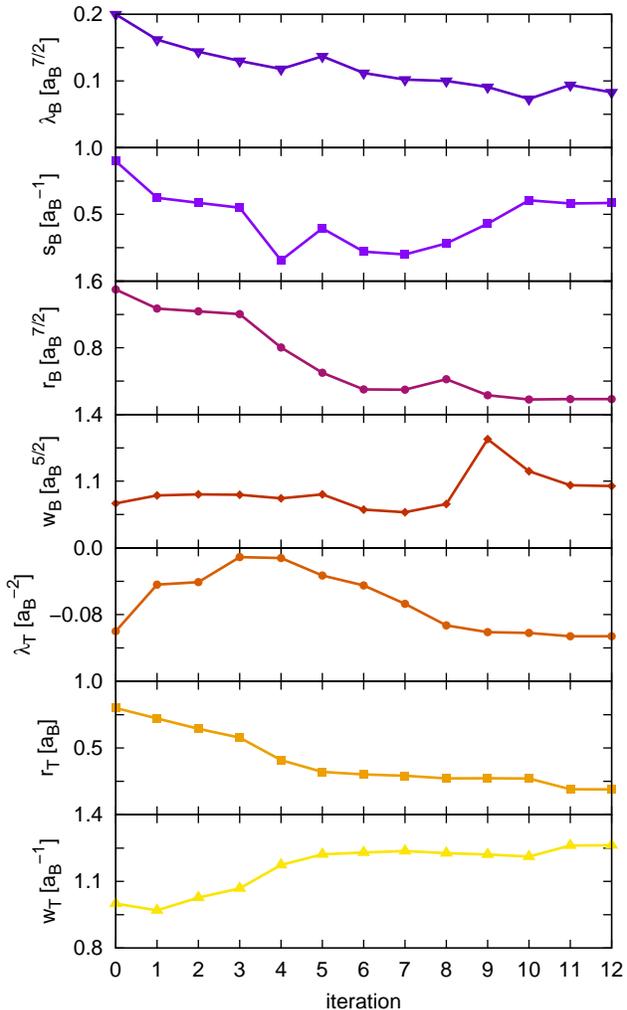}
\caption{(color online) Convergence of the backflow and three-body correlations of a 2D system
of $N=26$ electrons at $r_s =  1$ during their simultaneous optimization.}
\label{fig:12}
\end{figure}

The backflow correlations have been first optimized in absence of three-body correlations: 
the convergence of the VMC energy is illustrated in figure \eqref{fig:7}, and the flow in 
the parameter space impressed by the optimization algorithm in figure \eqref{fig:8}.
We obtain the energy $ -0.3846(2) \, Ry$, in good agreement with the value $-0.3839(4) \, Ry$ reported
in \cite{Kwon1996}.
%%%%%%%%%%%%%%%%%%%%%%%%%%%%%%%%%%%%%%%%%%%%%%%%%%%%%%%%%%%%%%%%%%%%%%%%%%%%%%%%%%%%%%%%%%%%%%%%%%%%%%%%%%%%%%%%%%%%%%%%%%%%%%%%
% ANDAMENTO ENERGIA HEG BF
% ANDAMENTO PARAMETRI HEG BF
%%%%%%%%%%%%%%%%%%%%%%%%%%%%%%%%%%%%%%%%%%%%%%%%%%%%%%%%%%%%%%%%%%%%%%%%%%%%%%%%%%%%%%%%%%%%%%%%%%%%%%%%%%%%%%%%%%%%%%%%%%%%%%%%
The gaussian factor has been subsequently optimized keeping the backflow correlations fixed at the parameter 
values  $\parm_{BF}=(\lambda_B,s_B,r_B,w_B)$ corresponding to the last step of figure \eqref{fig:7}. The 
convergence of the VMC energy is illustrated in figure \eqref{fig:9}, and the flow in the parameter space 
in figure \eqref{fig:10}.
%%%%%%%%%%%%%%%%%%%%%%%%%%%%%%%%%%%%%%%%%%%%%%%%%%%%%%%%%%%%%%%%%%%%%%%%%%%%%%%%%%%%%%%%%%%%%%%%%%%%%%%%%%%%%%%%%%%%%%%%%%%%%%%%
% ANDAMENTO ENERGIA HEG 3B
% ANDAMENTO PARAMETRI HEG 3B
%%%%%%%%%%%%%%%%%%%%%%%%%%%%%%%%%%%%%%%%%%%%%%%%%%%%%%%%%%%%%%%%%%%%%%%%%%%%%%%%%%%%%%%%%%%%%%%%%%%%%%%%%%%%%%%%%%%%%%%%%%%%%%%%
A simultaneous optimization of the three-body and backflow correlations has been finally carried out, starting from a randomly chosen parameter configuration,
leading to the results illustrated in figures \eqref{fig:11} and \eqref{fig:12}.
We remark that, although a more rapid convergence of the backflow parameters is attained in absence of
the three-body correlations, at least for the system under study, the algorithm proves able to 
simultaneously handle parameters with different orders of magniture and pertaining to different parts 
of the WF.

%%%%%%%%%%%%%%%%%%%%%%%%%%%%%%%%%%%%%%%%%%%%%%%%%%%%%%%%%%%%%%%%%%%%%%%%%%%%%%%%%%%%%%%%%%%%%%%%%%%%%%%%%%%%%%%%%%%%%%%%%%%%%%%%
% ANDAMENTO ENERGIA HEG TOTALE
% ANDAMENTO PARAMETRI HEG 3B
%%%%%%%%%%%%%%%%%%%%%%%%%%%%%%%%%%%%%%%%%%%%%%%%%%%%%%%%%%%%%%%%%%%%%%%%%%%%%%%%%%%%%%%%%%%%%%%%%%%%%%%%%%%%%%%%%%%%%%%%%%%%%%%%
\begin{table}
\centering
\begin{tabular}{c c}
\hline\hline \\[1ex]
N &   t  (sec)  \\[0.5ex]
\hline \\[0.5ex]
\,\,2   &  \,\,\,\, 0.15   \\[0.5ex]
\,10    &  \,\,\,\,\, 0.55   \\[0.5ex] 
\,26    &  \,\,\,\,   5.90 \\[0.5ex]
\,42    &  \,\,      22.50 \\[0.5ex]
\,58    &  \,\,      56.97 \\[0.5ex]
\,72    &  \,       108.00 \\[0.5ex]
\,98    &  \,   264.54  \\[0.5ex]
162 &       1141.35 \\[0.5ex]
242  &      3727.22 \\[0.5ex]
\hline\hline
\end{tabular}
 \caption{Duration per core $t$ (in seconds) of $100$ optimization blocks each made of $10$ VMC  steps
in for several numbers $N$ of electrons. The third degree polynomial $t(N) = a_2 N^2 + a_3 N^3$,
with $a_2 = 2.72226 \times 10^{-3} $ and $a_3 = 2.51738 \times 10^{-4}$ fits the data with reduced sum of square
residuals equal to $0.72$. Attempting to fit a quartic polynomial, adding a term $a_4 N^4$, we find
a coefficient $a_4 \sim 10^{-10}$ with negative sign, compatible with zero, confirming the scaling
of the algorithm.
}
\label{tabtimes}
\end{table}
In table \eqref{tabtimes} we provide estimates of the duration of optimization
runs, confirming the cubic scaling of the methodology in the number of particles.
The duration estimates were obtained using the WTIME function of the MPI library,
monitoring runs in which solely backflow correlations were optimized.
In the caption we show that the execution time per CPU actually scales as $N^3$,
a further confirmation of the key result of the present work, and a quantitative 
estimate of the performance of the algorithm.

%We remark that, even though the algorithm 

\section{Conclusions}
We have shown that, for correlated WFs containing two and three-body together
with backflow correlations, it is possible to implement the Linear Method to
optimize the variational parameters with a favorable complexity
$\mathcal{O}\left( N^3\right)$, 
$N$ being the number of particles.
We have described the algorithm in full detail showing the non-trivial
recipes to evaluate  the derivatives with respect to the variational parameters
and also the overlap and the energy matrices, attaining the best
possible complexity allowed by the need to perform VMC calculations, which already scale
as $N^3$.

%%%%%%%%%%%%%%%%%%%%%%%%%%%%%%%%%%%%%%%%%%%%%%%%%%%%%%%%%%%%%%%%%%%%%%%%%%%%%%%%%%%%%%%%%%%%%%%%%%%%%%%%%%%%%%%%%%%%%%%%%%%%%%%%
\section{Acknowledgments}
We acknowledge the CINECA and the Regione Lombardia
award, under the LISA initiative, for the availability of
high-performance computing resources and support.
One of the authors (M. M.) would like to acknowledge funding provided 
by the Dr. Davide Colosimo Award, 
celebrating the memory of physicist Davide Colosimo.

\appendix

\section{Efficient Optimization of Backflow Correlations}\label{appA}

In the present section, the quantities \eqref{bf3}, \eqref{bf4} and \eqref{bf7} will be computed and 
proved to have computational cost scaling as $\mathcal{O}( N^3 )$, $N$ being the number of particles 
constituting the system. The notations of reference \cite{Kwon1993}, Appendix B, will be adopted.
%%%%%%%%%%%%%%%%%%%%%%%%%%%%%%%%%%%%%%%%%%%%%%%%%%%%%%%%%%%%%%%%%%%%%%%%%%%%%%%%%%%%%%%%%
To this purpose, the following intermediate tensors need to be computed, with the numbers in brackets 
denoting the computational complexity:
\begin{enumerate}
\item the quasiparticle coordinates and their first and second derivatives [$N^2$]:
\begin{equation}
\label{int_tensors1}
\begin{split}
           x_{l\beta} &= r_{l\beta} + \sum_{j \neq l} \eta(r_{lj}) \left( r_{l\beta} - r_{j\beta} \right) \\
A^{\alpha \beta}_{il} &= \partial_{r_{i\alpha}} x_{l\beta} \quad 
B^{\alpha \beta}_{il}  = \partial^2_{r_{i\alpha}} x_{l\beta} \\
C^\beta_l             &= \partial_{p_j} x_{l\beta} = \sum_{j \neq l} \partial_{p_j}\eta(r_{lj}) \left( r_{l\beta} - r_{j\beta} \right) \\
D^{\alpha \beta}_{il} &= \partial_{p_j} A^{\alpha \beta}_{il} \quad
E^{\alpha \beta}_{il}  = \partial_{p_j} B^{\alpha \beta}_{il} \\
\end{split}
\end{equation}
\item the backflow matrix and its first, second and third derivatives [$N^2$]:
\begin{equation}
\label{int_tensors2}
\begin{split}
\varphi_{kl} &= \varphi_k(x_l) \\
\varphi^\beta_{kl}                                   &= \frac{ \partial   \varphi_{kl} }{ \partial_{x_{l\beta}}                              } \\
\varphi^{\beta\beta^\prime}_{kl}                     &= \frac{ \partial^2 \varphi_{kl} }{ \partial_{x_{l\beta}} \partial_{x_{l\beta^\prime}} } \\
\varphi^{\beta\beta^\prime\beta^{\prime\prime}}_{kl} &= \frac{ \partial^3 \varphi_{kl} }{ \partial_{x_{l\beta}} \partial_{x_{l\beta^\prime}} 
                                                                                          \partial_{x_{l\beta^{\prime\prime}}}} 
\end{split}
\end{equation}
\item the inverse $\varphi^{-1}$ of the backflow matrix [$N^3$] and the tensors [at most $N^3$]:
\begin{equation}
\label{int_tensors3}
\begin{split}
F^\beta_{kl}                      &= \sum_r        \varphi^{-1}_{kr} \varphi^\beta_{rl} \\
G_{kl}                            &= \sum_r        \varphi^{-1}_{kr} \partial_{p_j} \varphi_{rl} 
                                   = \sum_\beta F^\beta_{kl} C^\beta_l \\
%\sum_{r\beta} \varphi^{-1}_{kr} \varphi^\beta_{rl} C^\beta_l \\
H^{\beta\beta^\prime}_{ll^\prime} &= \sum_{i\alpha}               A^{\alpha \beta}_{il} A^{\alpha \beta^\prime}_{il^\prime} \\
J^{\beta\beta^\prime}_{kl}        &= \sum_r        \varphi^{-1}_{kr} \varphi^{\beta\beta^\prime}_{rl} \\
\end{split}
\end{equation}
\end{enumerate}
the tensor $G_{kl}$ has been explicited making use of the chain rule $\partial_{p_j}\varphi_{kl} = \sum_\beta 
\partial_{p_j}x_{l\beta} \varphi^\beta_{kl} = \sum_\beta C_l^\beta \varphi^\beta_{kl}$ and observing that:
\begin{equation}
G_{kl} = \sum_r \varphi^{-1}_{kr} \varphi^\beta_{rl} C_l^\beta = \sum_\beta F^\beta_{kl} C_l^\beta
\end{equation}
%%%%%%%%%%%%%%%%%%%%%%%%%%%%%%%%%%%%%%%%%%%%%%%%%%%%%%%%%%%%%%%%%%%%%%%%%%%%%%%%%%%%%%%%%
Recalling equations \eqref{bf3} and \eqref{int_tensors3}, we readily conclude that [$N$]:
\begin{equation}
\partial_{p_j} \mathcal{Z}(\parm,\cnf) = - \sum_{l} G_{ll}
\end{equation}
%%%%%%%%%%%%%%%%%%%%%%%%%%%%%%%%%%%%%%%%%%%%%%%%%%%%%%%%%%%%%%%%%%%%%%%%%%%%%%%%%%%%%%%%%
The quantity $\partial_{r_{i\alpha}} \partial_{p_j} \mathcal{Z}(\parm,\cnf)$ results from the difference of two terms:
\begin{enumerate}
\item $\mbox{tr}\left( \varphi^{-1} \partial_{r_{i \alpha}} \varphi \, \varphi^{-1} \partial_{p_j} \varphi \right)$
\item $\mbox{tr}\left( \varphi^{-1} (\partial_{r_{i \alpha}} \partial_{p_j} \varphi) \right)$
\end{enumerate}
which, recalling equations \eqref{bf4}, \eqref{int_tensors1} and the chain rule:
\begin{equation}
\begin{split}
\partial_{r_{i \alpha}}            \varphi_{kl} &= \sum_{l\beta}                      A^{\alpha\beta}_{il} \partial_{x_{l\beta}} \varphi_{kl} \\
\partial_{r_{i \alpha}} \partial_p \varphi_{kl} &= \sum_\beta                         D^{\alpha\beta}_{il}        \varphi^\beta_{kl} 
                                                + \sum_{\beta\beta^\prime} C^\beta_l A^{\alpha\beta^\prime}_{il} \varphi^{\beta\beta^\prime}_{kl}
\end{split}
\end{equation}
can be cast in the form [$N^2$]:
\begin{equation}
  \mbox{tr}\left( \varphi^{-1} \partial_{r_{i \alpha}} \varphi \, \varphi^{-1} \partial_{p_j} \varphi \right) = 
  \sum_{lk\beta} A^{\alpha \beta}_{il} G_{lk} F^\beta_{kl} \\
\end{equation}
\begin{equation}
\mbox{tr}\left( \varphi^{-1} (\partial_{r_{i \alpha}} \partial_{p_j} \varphi) \right) = 
\sum_{l\beta\beta^\prime} A^{\alpha \beta}_{il} C_l^{\beta^\prime} J^{\beta\beta^\prime}_{ll} +
\sum_{l\beta} D^{\alpha \beta}_{il} F^\beta_{ll} \\ 
\end{equation}
%%%%%%%%%%%%%%%%%%%%%%%%%%%%%%%%%%%%%%%%%%%%%%%%%%%%%%%%%%%%%%%%%%%%%%%%%%%%%%%%%%%%%%%%%
The quantity $\partial^2_{r_{i \alpha}} \partial_{p_j} Z(\parm,\cnf)$ results from a linear combination of the terms:
\begin{enumerate}
\item $\mbox{tr}\Big((\varphi^{-1}\partial^2_{r_{i \alpha}} \varphi) \varphi^{-1} \partial_{p_j} \varphi \Big)$
\item $\mbox{tr}\Big((\varphi^{-1}\partial_{r_{i \alpha}}   \varphi)(\varphi^{-1} \partial_{r_{i \alpha}} \varphi)
                                                                    (\varphi^{-1} \partial_{p_j} \varphi) \Big)$
\item $\mbox{tr}\Big((\varphi^{-1}\partial_{r_{i \alpha}}   \varphi)(\varphi^{-1} \partial_{r_{i \alpha}} \partial_{p_j} \varphi) \Big)$
\item $\mbox{tr}\Big(  \varphi^{-1}\partial^2_{r_{i \alpha}} \partial_{p_j} \varphi \Big)$
\end{enumerate}
Recalling $\partial^2_{r_{i \alpha}}\varphi_{kl}=\sum_\beta               B^{\alpha\beta}_{il}\varphi^\beta_{kl}
          +\sum_{\beta\beta^\prime} A^{\alpha\beta}_{il} A^{\alpha\beta^\prime}_{il} \varphi^{\beta\beta^\prime}_{kl}$
the first term can be cast in the form [$N^3$]:
\begin{equation}
\begin{split}
  & \sum_{i\alpha} \mbox{tr}\Big((\varphi^{-1}\partial^2_{r_{i \alpha}}\varphi)\varphi^{-1}\partial_{p_j}\varphi\Big) = \\
= & \sum_\beta               \sum_{kl} \left( \sum_{i \alpha} B^{\alpha\beta}_{il} \right) F^\beta_{kl} G_{lk} + 
    \sum_{\beta\beta^\prime} \sum_{kl} H^{\beta\beta^\prime}_{ll} J^{\beta\beta^\prime}_{kl} G_{lk} \\
\end{split}
\end{equation}
The second term in the form [$N^3$]:
\begin{equation}
\begin{split}
  &\sum_{i\alpha} \mbox{tr}\Big((\varphi^{-1}\partial_{r_{i \alpha}}\varphi)(\varphi^{-1}\partial_{r_{i \alpha}}\varphi)(\varphi^{-1}\partial_{p_j}\varphi)\Big)=\\
= &\sum_{\beta\beta^\prime} \sum_{ll^\prime} H^{\beta\beta^\prime}_{ll^\prime} 
\left( \sum_k F^\beta_{kl} G_{l^\prime k} \right) F^{\beta^\prime}_{ll^\prime}
\end{split}
\end{equation}
The third term in the form [$N^3$]:
\begin{equation}
\begin{split}
   &\sum_{i\alpha}   \mbox{tr}\Big((\varphi^{-1}\partial_{r_{i \alpha}}\varphi)(\varphi^{-1}\partial_{r_{i \alpha}}\partial_{p_j}\varphi)\Big) = \\
 = &\sum_{ll^\prime} \sum_{\beta\beta^\prime} H^{\beta\beta^\prime}_{ll^\prime}                F^{\beta^\prime}_{l^\prime l} 
                                              J^{\beta^\prime\beta^{\prime\prime}}_{ll^\prime} C_{l^\prime}^{\beta^{\prime\prime}} 
  + \sum_{kli} \sum_{\alpha\beta\beta^\prime} F^{\beta}_{lk} A^{\alpha\beta}_{ik} D^{\alpha\beta^\prime}_{il} F^{\beta^\prime}_{kl}
\end{split}
\end{equation}
Finally, recalling that:
\begin{equation}
\begin{split}
  &\partial^2_{r_{i\alpha}} \partial_{p_j} \varphi_{kl} = 
   \sum_\beta E^{\alpha\beta}_{il} \varphi^\beta_{kl} + \sum_{\beta\beta^\prime} B^{\alpha\beta}_{il} C^{\beta^\prime}_l \varphi^{\beta\beta^\prime}_{kl} \\
+ &\sum_{\beta\beta^\prime\beta^{\prime\prime}} A^{\alpha\beta}_{il} A^{\alpha\beta^\prime}_{il} C^{\beta^{\prime\prime}}_l 
   \varphi^{\beta\beta^\prime \beta^{\prime\prime}}_{kl} 
+
2  \sum_{\beta\beta^\prime} D^{\alpha\beta}_{il} A^{\alpha\beta^\prime}_{il} \varphi^{\beta\beta^\prime}_{kl}
\end{split}
\end{equation}
the fourth term can be cast in the form [$N^2$]:
\begin{equation}
\begin{split}
  & \sum_{i\alpha} \mbox{tr}\Big(\varphi^{-1}(\partial^2_{r_{i \alpha}}\partial_{p_j}\varphi)\Big) = 
    \sum_{l\beta} \big( \sum_{i\alpha} E^{\alpha\beta}_{il} \big) F^\beta_{ll} \\
+ & 2 \sum_{l\beta\beta^\prime} \Big( \sum_{i\alpha} D^{\alpha\beta}_{il} A^{\alpha\beta^\prime}_{il} \Big) J^{\beta\beta^\prime}_{ll} 
+     \sum_{l\beta\beta^\prime} \big( \sum_{i\alpha} B^{\alpha\beta}_{il} \big) C^{\beta^\prime}_l          J^{\beta\beta^\prime}_{ll} \\ 
+ & \sum_{l\beta\beta^\prime\beta^{\prime\prime}}    H^{\beta\beta^\prime}_{ll} C^{\beta^{\prime\prime}}_l
 \big( \sum_k \varphi^{-1}_{lk} \varphi^{\beta\beta^\prime\beta^{\prime\prime}}_{kl} \big)
\end{split}
\end{equation}
The recommendations outlined in the present Appendix are to be respected in order to contain the computational cost of the optimization 
procedure.
Further simplifications in the calculation of the intermediate tensors \eqref{int_tensors2} are possible in homogeneous systems, where
the backflow orbitals are plane waves $\varphi_{il} = e^{i \vett{k}_i \cdot \vett{x}_l}$.

\section{Complex-valued WFs} \label{appB}
In this appendix we present the generalization of the
optimization algorithm to complex-valued WFs.
In the case of Slater determinants of plane waves, we observe that,
denoting by $*$ the complex conjugation:

\begin{equation}
\det\left( e^{i     \vett{k}_i \cdot \vett{x}_l} \right)^*
= \det\left( e^{- i \vett{k}_i \cdot \vett{x}_l} \right)
\end{equation}
In the study of the ground state in periodic boundary conditions a set of $\vett{k}$-points closed under the time-reversal operation 
$\vett{k}\to -\vett{k}$ ensures the reality of the wave-function. On the other hand, different choices of boundary conditions or the 
study of some particular excited states or the presence of an external magnetic field require the formalism of complex-valued WFs.
The function $\mathcal{E}(\parm)$ to be optimized with respect to pararameter variations $\Delta\parm$ takes the form:
\begin{equation}
\mathcal{E}(\parm) = 
\frac{ \begin{pmatrix} 1 \,\,  \Delta \parm^T \end{pmatrix}
       \begin{pmatrix} \mathcal{E}(\parm_0) & \vett{g}^T \\
                       \vett{g}^*           & \overline{\mathcal{H}} \\
       \end{pmatrix}
       \begin{pmatrix} 1 \\ \Delta \parm \end{pmatrix} }
     { \begin{pmatrix} 1 \,\,  \Delta \parm^T \end{pmatrix}
       \begin{pmatrix} 1                    & \vett{s}^T \\
                       \vett{s}^*           & \overline{\mathcal{S}} \\
       \end{pmatrix}
       \begin{pmatrix} 1 \\ \Delta \parm \end{pmatrix} }
\end{equation}
where $\mathcal{E}(\parm_0)$ is the current value of the energy,
\begin{equation}
g_j=\frac{ \braket{\overline{\Psi}_0|\hat{H}|\overline{\Psi}_j} }
          { \braket{\overline{\Psi}_0|\overline{\Psi}_0} }
\end{equation}
satisfying:

\begin{equation}
\partial_{p_j} \mathcal{E}(\parm_0) = g_j + g^*_j, \quad 
\end{equation}
and
\begin{equation}
s_j=\frac{ \braket{\overline{\Psi}_0|\overline{\Psi}_j} }
                     { \braket{\overline{\Psi}_0|\overline{\Psi}_0} }
\end{equation}
The overlap and energy matrix are defined exactly as
in the real case:

\begin{equation}
\overline{\mathcal{S}}_{ij} = \frac{ \braket{\overline{\Psi}_i|\overline{\Psi}_j} }
                                    { \braket{\overline{\Psi}_0|\overline{\Psi}_0} }
, \quad \overline{\mathcal{H}}_{ij} = \frac{ \braket{\overline{\Psi}_i|\hat{H}|\overline{\Psi}_j} }
                                        { \braket{\overline{\Psi}_0|\overline{\Psi}_0} }
\end{equation} 
The generalized eigenvalue problem on which the method relies,
in the complex case, is:

\begin{equation}
       \begin{pmatrix} \mathcal{E}(\parm_0) & \vett{g}^T \\
                       \vett{g}^*           & \overline{\mathcal{H}} \\
       \end{pmatrix}
       \begin{pmatrix} 1 \\ \Delta \parm \end{pmatrix} 
       =
       \mathcal{E}
       \begin{pmatrix} 1                    & \vett{s}^T \\
                       \vett{s}^*           & \overline{\mathcal{S}} \\
       \end{pmatrix}
       \begin{pmatrix} 1 \\ \Delta \parm \end{pmatrix}
\end{equation}
Except for the care to take the complex conjugation whenever necessary, the most 
noticeable difference with respect to the real case is the presence of the vector 
$\vett{s}$, which has to be estimated.
In the complex case, in fact, the normalization constraint $0 = \partial_{p_j} 
\braket{ \tilde{\Psi}(\parm) | \tilde{\Psi}(\parm)}$
implies that the overlap between $\ket{\overline{\Psi}_0}$ and $\ket{\overline{\Psi}_i}$
is a non-vanishing purely imaginary number.
The steps of the linear method, then, proceed exactly as in the real case.

\bibliographystyle{unsrt}

\begin{thebibliography}{9}
\bibitem{Toulouse2007}    J. Toulouse and C.J. Umrigar,
                          \emph{J. Chem. Phys.} 126, 084102 (2007)
\bibitem{Szabo1996}       For a comprehensive review of the existing numerical wave-function-based
                          methodologies see for example the book:
                          A. Szabo and N.S. Ostlund,
                          {\emph{Modern Quantum Chemistry: Introduction to Advanced Electronic 
                          Structure Theory}}, Dover Publications (1996)               
\bibitem{Whitlock1986}    For a comprehensive review of Quantum Monte Carlo methods see for example the book: 
                          M.H. Kalos and P. A. Whitlock {\emph{Quantum Monte Carlo}}, in {\emph{Monte Carlo 
                          Methods}}, Wiley (1986). 
\bibitem{gfmc}            M. H. Kalos, {\em Phys. Rev.} { \bf 128}, 1891 (1962)
\bibitem{rept}            S. Baroni and S. Moroni,             {\em Phys. Rev. Lett.} {\bf 82}, 4745 (1999)
\bibitem{pigs}            A. Sarsa, K.E. Schmidt and W. Magro, {\em J. Chem. Phys.} {\bf 113}, 1366 (2000)
\bibitem{spigs}           D. E. Galli and L. Reatto,          {\em Mol. Phys.} {\bf 101}, 1697 (2003).
\bibitem{patate}          M. Rossi, M. Nava, L. Reatto, and D.E. Galli, {\em J. Chem. Phys.} {\bf 131}, 154108 (2009).
\bibitem{davide}          D. E. Galli, E. Cecchetti, and L. Reatto, {\em Phys. Rev. Lett.}, {\bf 77}, 5401 (1996).
\bibitem{ettore}          E. Vitali, P. Arrighetti, M. Rossi, and D.E. Galli, {\em{Mol. Phys.}}, {\bf 109}, 2855 (2011).
\bibitem{Feynman1965}     R. P. Feynman and A. R. Hibbs,  
                          {\em Quantum Mechanics and Path Integrals}, McGraw-Hill (1965)
\bibitem{Caffarel1988}    M. Caffarel and P. Claverie, 
                          \emph{J. Chem. Phys.} 88, 108 (1988)
\bibitem{Caffarel1989}    M. Caffarel,
                          \emph{Stochastic methods in quantum mechanics} in
                          \emph{Numerical Determination of the Electronic Structure of Atoms, Diatomic and Polyatomic                          Molecules},  Kluwer Academic Publishers (1989)
\bibitem{Holz2003}        M. Holzmann, D. M. Ceperley, C. Pierleoni and K. Esler, 
                          \emph{Phys. Rev. E} 68, 046707 (2003)
\bibitem{Umrigar1988}     C. J. Umrigar, K. G. Wilson, and J. W. Wilkins
                          \emph{Phys. Rev. Lett.} 60, 1719 (1988)                          
\bibitem{Levenberg1944}   K. Levenberg,
                          \emph{Quart. Appl. Math.} 2, 164 (1944)
\bibitem{Marquardt1963}   D. Marquardt,
                          \emph{SIAM J. Appl. Math.} 11, 431 (1963)
\bibitem{Kent1999}        P.R.C. Kent, R.J. Needs and G. Rajagopal,
                          \emph{Phys. Rev. B} 59, 12344 (1999)
\bibitem{Rappe2000}       X. Lin, H. Zhang, and A. M. Rappe,
                          \emph{J. Chem. Phys.} 112, 2650 (2000)
\bibitem{Rappe2005}       M. W. Lee, M. Mella, and A. M. Rappe,
                          \emph{J. Chem. Phys.} 112, 244103 (2005)
\bibitem{Filippi2005}     C.J. Umrigar and Claudia Filippi,
                          \emph{Phys. Rev. Lett.} 94, 150201 (2005)
\bibitem{Sorella2005}     S. Sorella,
                          \emph{Phys. Rev. B (Rapid Comm.)} 71, 241103 (2005)
\bibitem{Umrigar2007}     C.J. Umrigar, J. Toulouse, C. Filippi, S. Sorella and R. G. Hennig,
                          \emph{Phys. Rev. Lett.} 98, 110201 (2007)
\bibitem{Huang1990}       S. Huang, Z. Sun, and W. A. Lester Jr.
                          \emph{J. Chem. Phys.} 92, 597 (1990)
\bibitem{Nightingale2001} M. P. Nightingale and V. Melik-Alaverdian,
                          \emph{Phys. Rev. Lett.} 87, 043401 (2001)
\bibitem{sorella2010}     S. Sorella and L. Capriotti, \emph{J. Chem. Phys.} 133, 234111 (2010)
\bibitem{nota_eigen}           We notice that, in principle, the generalized eigenvalue equation \eqref{lm8}
                          might admit solutions whose first component is zero; equation \eqref{lm7} shows
                          that such solution corresponds to parameter variations that are orthogonal
                          to the energy gradient and thus remain tangent to the hypersurface of
                          constant energy, bringing no improvements.
\bibitem{Umrigar2008}     J. Toulouse and C.J. Umrigar,
                          \emph{J. Chem. Phys.} 128, 174101 (2008)
\bibitem{Sorella2001}     S. Sorella,
                          \emph{Phys. Rev. B} 64, 024512 (2001)
\bibitem{Tikhonov1977}    A.N. Tikhonov and V. Y. Arsenin,
                          \emph{Solution of Ill-posed Problems}, Winston \& Sons (1977)
\bibitem{Kalos1980}       K.E. Schmidt, M.W. Kalos, M.A. Lee and G.V. Chester,
                          \emph{Phys. Rev. Lett.} 45, 573 (1980)
\bibitem{Saverio1995}     S. Moroni, S. Fantoni and G. Senatore,
                          \emph{Phys. Rev. B} 52, 13547 (1995)
\bibitem{Kalos1981}       K.E. Schmidt, M.A. Lee, M.W. Kalos and G.V. Chester,
                          \emph{Phys. Rev. Lett.} 47, 807 (1981)
\bibitem{Carlson1989}     R. M. Panoff and J. Carlson,
                          \emph{Phys. Rev. Lett.} 62, 1130 (1989)
\bibitem{Kwon1993}        Y. Kwon, D. M. Ceperley and R. M. Martin,
                          \emph{Phys. Rev. B} 48, 12037 (1993)
\bibitem{Lopez2006}       P. Lopez-Rios, A. Ma, N. D. Drummond, M. D. Towler, R. J. Needs
                          \emph{Phys. Rev. E} 74, 066701 (2006)
\bibitem{Ceperley1991}    D. M. Ceperley, 
                          \emph{J. Stat. Phys.} 63, 1237 (1991)
\bibitem{Foulkes2001}     W. M. C. Foulkes, L. Mitas, R. J. Needs, and G. Rajagopal,
                          \emph{Rev. Mod. Phys.} 73, 33 (2001)
\bibitem{Magnus1999}      J. R. Magnus and N. Neudecker, 
                          \emph{Matrix Differential Calculus with Applications in Statistics and Econometrics}, 
                          Wiley (1999)
\bibitem{Aziz1979}        R. A. Aziz, V. P. S. Nain, J. S. Carley, W. L. Taylor and G. T. McConville
                          \emph{J. Chem. Phys.} 70, 4330 (1979)
\bibitem{McMillan1964}    W. L. McMillan,
                          \emph{Phys. Rev.} 138, A422 (1964)
\bibitem{Gaskell1961}     T. Gaskell, 
                          \emph{Proc. Phys. Soc.} 77, 1182 (1961)
\bibitem{Gaskell1962}     T. Gaskell, 
                          \emph{Proc. Phys. Soc.} 80, 1091 (1962)
\bibitem{Ceperley1989}    B. Tanatar and D. M. Ceperley, 
                          \emph{Phys. Rev. B} 39, 5005 (1989)
\bibitem{Ceperley1996}    K. E. Schmidt and D. M. Ceperley,
                          \emph{Monte Carlo Techniques for Quantum Fluids, Solids and Droplets}
                          in \emph{The Monte Carlo Method in Condensed Matter Physics}, Springer-Verlag, (1992)
\bibitem{Kwon1996}        Y. Kwon, D. M. Ceperley and R. M. Martin, \emph{Phys. Rev. B} 53, 7376 (1996)
\end{thebibliography}

%\end{multicols}{2}

\end{document}